\documentclass[showpacs
,showkeys
,nofootinbib
,amsfonts
,floatfix
,twocolumn
,tightenlines
,aps
,prc
,longbibliography
,superscriptaddress
]{revtex4-2}
\pdfoutput=1

\usepackage{graphicx}  
\usepackage{mathtools}
\usepackage{amsmath, amssymb, bm,bbm}   
\usepackage{xcolor}
\usepackage{natbib}
\usepackage{hyperref}

\usepackage{longtable}
\usepackage{csvsimple} 
\usepackage{adjustbox} 

\usepackage{ifthen}

\usepackage{siunitx} 
\DeclareSIUnit\barn{b}

\usepackage{balance}

\usepackage{orcidlink}

\newcommand*\xbar[1]{%
   \hbox{%
     \vbox{%
       \hrule height 0.5pt 
       \kern0.3ex
       \hbox{%
         \kern-0.1em
         \ensuremath{#1}%
         \kern-0.1em
       }%
     }%
   }%
} 

\def\HeAlpha{$^3\mathrm{He}(\alpha,\gamma)^7\mathrm{Be}$}

\def\HeAlphaHead{$\bm{^3}$H\lowercase{e}$\bm{(\alpha,\gamma)^7}$B\lowercase{e}}

\def\P{\left(\!\frac{\stackrel{\rightarrow}{\nabla}}{m_\phi}
\!-\!\frac{\stackrel{\leftarrow}{\nabla}}{m_\psi}\!\right)}

\begin{document}

\title{Precision calculation of \texorpdfstring{\HeAlphaHead}{He3-Alpha}  for solar physics}

\author{%
 Ratna Khadka\,\orcidlink{0009-0009-6628-7028}}
\email{rk973@.msstate.edu}
\affiliation{Department of Physics \& Astronomy and HPC$^2$ Center for 
Computational Sciences, Mississippi State
University, Mississippi State, MS 39762, USA}

\author{%
Ling Gan\,\orcidlink{0009-0007-0263-7602}}
\email{linggan@arizona.edu Current address: Department of Physics, University of Arizona, Tucson, AZ 85721}
\affiliation{Department of Physics \& Astronomy and HPC$^2$ Center for 
Computational Sciences, Mississippi State
University, Mississippi State, MS 39762, USA}

\author{%
Renato Higa\, \orcidlink{0000-0002-6298-8128}}
\email{higa@if.usp.br}
\affiliation{Instituto de F\'isica, Universidade de S\~ao Paulo, R. do Mat\~ao Nr.1371, 05508-090, S\~ao Paulo, SP, Brazil}

\author{%
Gautam Rupak\,\orcidlink{0000-0001-6683-177X}}
\email{grupak@ccs.msstate.edu}
\affiliation{Department of Physics \& Astronomy and HPC$^2$ Center for 
Computational Sciences, Mississippi State
University, Mississippi State, MS 39762, USA}

\begin{abstract} 
We calculate the cross section for radiative capture \HeAlpha ~at next-to-next-to-leading order (NNLO). At this order of perturbation,  momentum dependent two-body currents make their first appearance. We provide a model-independent construction of these currents from gauge and Galilean invariance, where the general framework for constructing higher-order two-body currents in low-energy effective field theories becomes evident. The \HeAlpha ~astrophysical S-factor $S_{34}(0)= 0.564^{+0.17}_{-0.015}~\si{\kilo\eV\barn}$ is obtained from a Bayesian analysis at NNLO, with an additional nominal theoretical uncertainty $\SI{\pm0.017}{\kilo\eV\barn}$ of 3\%. 
 \end{abstract}

\maketitle

\section{Introduction}
\label{sec:Introduction}

Neutrinos, due to their frail interaction with ordinary matter, are 
excellent probes of nuclear reactions taking place in the interior of 
the Sun~\cite{Acharya:2024lke}. Reaction rates of these nuclear processes determine the 
chemical composition, age, and evolution of the Sun and similar stars 
according to the successful standard solar model (SSM)~\cite{PhysRevLett.78.171,Bahcall:2000nu,Bahcall_2005}. 
For instance, neutrinos from the $pp$ chain and CNO cycle determine 
the temperature at the core of the Sun that agrees with helioseismological 
observations to a precision better than 0.2\%~\cite{PhysRevLett.78.171}.

The confirmation of neutrino oscillations~\cite{Fukuda:2001nj,Ahmad:2001an} heralded  
neutrino physics and related areas into the precision era. 
Neutrino oscillations demand physics beyond the standard model of particle 
physics, with questions like mass hierarchy, mixing angles, CP phases, 
and matter oscillations 
calling for answers. Solar neutrinos played a pivotal role in confirming 
such oscillations while establishing the SSM on solid grounds. 
However, to address the current open questions in neutrino physics 
the uncertainties in the predicted solar neutrino fluxes need better 
precision than the current $7$-$15$\%~\cite{Acharya:2024lke}. An important source of 
uncertainty comes from reaction rates of certain nuclear processes in 
the $pp$ chain. Among them, the 
$^3\text{He}(\alpha,\gamma)^7\text{Be}$ plays a central role. 
The astrophysical S-factor of this reaction, $S_{34}$, determines the 
flux of neutrinos from the $\beta^{+}$ decay of $^8\text{B}$ and from 
the electron capture on $^7\text{Be}$ ---they are proportional to 
$[S_{34}]^{0.81}$ and $[S_{34}]^{0.86}$, respectively~\cite{Cyburt:2008up,Bahcall:1987jc}. To meet the demands of future neutrino detectors like Hyper-K~\cite{Hyper-Kamiokande:2018ofw}, 
SNO+~\cite{PhysRevD.99.012012}, 
JUNO~\cite{Abusleme_2021},
and DUNE~\cite{PhysRevLett.123.131803}, 
the main source of uncertainty of $^7\text{Be}$ and $^8\text{B}$ 
neutrino fluxes, from $S_{34}(0)$, needs to be known with improved 
precision~\cite{Cyburt:2008up,PhysRevD.69.123519}.

The decadal publication Solar Fusion III (SF3)  recently made recommendation for the \HeAlpha ~S-factor $S_{34}(0)=\SI{0.561\pm0.18\pm0.022}{\kilo\eV\barn}$~\cite{Acharya:2024lke}. The first set of uncertainty is from the Bayesian fits to data, and the second set is an estimate of the theoretical uncertainty arrived at from multiple data analysis with several theoretical calculations: Halo EFT~\cite{Higa:2016igc,Premarathna:2019tup}, Halo EFT~\cite{Zhang_2020} and R matrix BRICK~\cite{Odell:2021nmp}.  The final SF3 recommendations were based on the next-to-leading order (NLO) calculations performed in Halo EFT from Ref.~\cite{Higa:2016igc}. Nominally, for such a calculation, one would adopt a 10\% theoretical uncertainty from higher-order corrections. These corrections affect the asymptotic normalization constant (ANC) which is momentum independent, initial state interaction that is suppressed at low momentum, and two-body currents that would also be suppressed at low momentum. Thus, it is reasonable that the theoretical uncertainty is smaller than the conventional 10\% estimate once the ANC is fitted to data as long as the NLO calculation is able to describe the features in the data as a function of momentum. Nevertheless, it is necessary to confirm this expectation with an explicit next-to-next-to-leading order (NNLO) calculation. It  is accomplished in this paper.

\section{Low-Momentum Cross Section}
\label{sec:farmalism}

In  halo EFT, the $^3$He and $\alpha$ nuclei are treated as point-like particles, and the ground and excited states of $^7$Be are considered two-body bound states of  $^3$He and $\alpha$ point-like nuclei. The ground state of $^7$Be has spin-parity assignment $\frac{3}{2}^{-}$ with binding energy $B_0=\SI{1.5866}{\mega\eV}$, and the excited state has spin-parity $\frac{1}{2}^{-}$ with binding energy $B_1=\SI{1.1575}{\mega\eV}$.

The spin-parity assignments $\frac{1}{2}^+$ and $0^+$, respectively, for the $^3$He and $\alpha$ nuclei imply the ground and excited states of $^7$Be are $p$-wave bound states. Capture at low energy is dominated by E1 transition that can proceed from both initial  $s$- and $d$-wave states. The Gamow energy peak for this reaction is at $\sim\SI{23}{\kilo\eV}$~\cite{Takacs:2015}. The transition  amplitude  at center-of-mass (cm) momentum $p$ can be written as~\cite{Higa:2016igc}: 
\begin{multline}
    i\mathcal A^{(\zeta)}(p) \equiv \sum_{a,b=1}^3
    \bigg[S^{(\zeta)}(p)\delta_{ab} \\
    +3D^{(\zeta)}(p) (\hat{p}_a\hat{p}_b-\frac{1}{3}\delta_{ab})\bigg]\Gamma_{ab}\,,
\end{multline}
where $\zeta=0,1$ correspond to capture to the $^2P_{3/2}$ ground and $^2P_{1/2}$ excited states of $^{7}$Be, respectively, in the spectroscopic notation $^{2S+1}L_J$ for total spin $S$, orbital angular momentum $L$ and angular momentum $J$. $S^{(\zeta)}$ is associated with the $s$-wave capture and $D^{(\zeta)}$ is associated with the $d$-wave capture. The spinor structure for projection to the final $p$-wave bound state is ~\cite{Higa:2016igc}
\begin{multline}\label{eq:spinor}
\Gamma^{(\zeta)}_{ab}=\left(\frac{e Z_\phi}{m_\phi}
-\frac{e Z_\psi}{m_\psi}\right)\sqrt{\frac{6\pi}{\mu}}\sqrt{\mathcal Z^{(\zeta)}}
\sqrt{2m_\phi}\\
\times \sum_{\alpha,\beta=1}^2\sum_{i=1}^3\epsilon_a^\ast(\bm{k}) {U_{i}^{\ast\alpha,\zeta}}
(-{\bm k}) P_{ib}^{\alpha\beta,\zeta}U^{\beta,\psi}(-{\bm p})\,,
\end{multline}
where $\bm{\epsilon}(\bm{k})$ is the outgoing  photon polarization with cm momentum $\bm{k}$; $U^{\beta, \psi}(-\bm{p})$ is the spinor field for the incoming spin-$\frac{1}{2}$ $^3$He nucleus with cm momentum $-\bm{p}$ and spin index $\beta$, mass $m_\psi=\SI{2808.4}{\mega\eV}$ and charge  $Z_\psi=2$;  $m_\phi=\SI{3727.4}{\mega\eV}$ is the mass and  $Z_\phi=2$ is the charge for the incoming spin-$0$ $\alpha$ nucleus with cm momentum $\bm{p}$;
$U^{\alpha,\zeta}_i(-\bm{k})$ is the spinor field for the $p$-wave final state with cm momentum $-\bm{k}$, spin index $\beta$, vector index $i$ and mass $M=m_\psi+m_\phi$. 
$\mu=m_\psi m_\phi/M$ is the reduced mass. The charges are in units of the proton charge $e$.   $\mathcal Z^{(\zeta)}$ is the wave function renormalization constant that we discuss later. The spinor for the final state carries both spin and vector index that are projected onto the relevant $p$-wave states with the projectors
\begin{align}
&P_{ij}^{\alpha\beta,\zeta}= \delta_{ij}\delta^{\alpha\beta}-\frac{1}{3}(\sigma_i\sigma_j)^{\alpha\beta} &\text{for } \zeta=0\,,\nonumber\\
&P_{ij}^{\alpha\beta,\zeta}= \frac{1}{3}(\sigma_i\sigma_j)^{\alpha\beta} &\text{for } \zeta=1\,.
\end{align}

The spinor fields in Eq.~(\ref{eq:spinor})   satisfy the completeness 
relations~\cite{Higa:2016igc}, when summed over the polarizations,  
\begin{align}
\sum_\text{pol.}U^{\alpha,\psi}(\bm{p})[U^{\beta,\psi}]^\ast(\bm{p})=2m_\psi\delta^{\alpha\beta}\,,\nonumber\\
\sum_\text{pol.}U^{\alpha,\zeta}_i(\bm{p})[U^{\beta,\zeta}_j]^\ast(\bm{p})=2MP_{ij}^{\alpha\beta, \zeta}\,.
\end{align}
This results in a  total cross section averaged over initial and summed over final spins: 
\begin{align}
\sigma(p) = \frac{1}{16\pi M^2}\frac{1}{2}\sum_{\zeta=0}^{1}
\frac{p^2+\gamma_\zeta^2}{2\mu p}|\mathcal M^{(\zeta)}(p)|^2 \,,
\end{align}
where
\begin{multline}
  |\mathcal M^{(\zeta)}(p)|^2=(2j+1)
  \left(\frac{ Z_\phi m_\psi}{M}-\frac{ Z_\psi m_\phi}{M}\right)^2\\
\times  \left[ |S^{(\zeta)}(p)|^2 +2|D^{(\zeta)}(p)|^2\right]  \frac{64\pi\alpha_\text{EM}  M^2 }{\mu} \frac{6\pi}{\mu}\mathcal Z^{(\zeta)}\, ,
\end{multline}
with $j=3/2$, binding momentum $\gamma_0=\sqrt{2\mu B_0}$ for the ground state $\zeta=0$; and $j=1/2$, binding momentum $\gamma_1=\sqrt{2\mu B_1}$ for the excited state $\zeta=1$ of $^7$Be, respectively. $\alpha_\text{EM}=e^2/(4\pi)=1/137$ is the fine structure constant. 

The cross section is used to define the astrophysical S-factor at cm energy $E=p^2/(2\mu)$
\begin{align}\label{eq:S34}
S_{34}(E)= E e^{2\pi\eta_p}\sigma(p=\sqrt{2\mu E})\,,
\end{align}
where the Sommerfeld parameter $\eta_p\!=\!{k_C}/{p}$ with the inverse Bohr radius $k_C \!=\! \alpha_\text{EM} Z_\psi Z_\phi\mu$. The branching ratio of capture to the excited over the ground state of $^7$Be is 
\begin{align}\label{eq:BR}
R(p) =\frac{p^2+\gamma_1^2}{p^2+\gamma_0^2} \frac{|\mathcal M^{(1)}(p)|^2}{|\mathcal M^{(0)}(p)|^2}\,.
\end{align}

The \HeAlpha ~cross section was calculated to NLO in a perturbative $Q/\Lambda$ expansion in Ref.~\cite{Higa:2016igc,Premarathna:2019tup}. The binding momenta 
$\gamma_0$, $\gamma_1$ along with the cm low momentum $p\lesssim \gamma_1\sim\gamma_0$ constitutes the small momentum scale $Q\sim \SI{70}{\mega\eV}$. The excited states  of $\alpha$, proton separation energy of $^3$He, pion physics, etc., define a high momentum scale $\Lambda\gtrsim 150$-$\SI{200}{\mega\eV}$ that is beyond the range of applicability of the halo EFT. 
We present the NNLO contribution below.

\section{Perturbation at NNLO}
\label{sec:perturbation}

The capture cross section contributions can be divided into three parts, and considering these separately is convenient for extending the calculation to NNLO. Schematically, the capture amplitude is $\langle\psi_\text{bound}|\operatorname{O}_\text{E1}|\psi_\text{in}\rangle$. 

Elastic scattering in the initial state $|\psi_\text{in}\rangle$ in the presence of short-ranged strong interaction and long-ranged Coulomb can be parameterized with the modified effective range expansion (ERE) for the Coulomb-subtracted phase shift $\delta_l$ in the $l$-th partial wave:
\begin{multline}\label{eq:ERE}
\left[\frac{\Gamma(2l+2)}{2^l\Gamma(l+1)}\right]^2 [C_l(\eta_p)]^2 p^{2l+1}
(\cot\delta_l-i)=-\frac{1}{a_l}+\frac{1}{2} r_l p^2\\
+\frac{1}{4}s_l p^4+\cdots
-\frac{2k_C\, p^{2l}}{\Gamma(l+1)^2}
\frac{|\Gamma(l+1+i\eta_p)|^2}{|\Gamma(1+i\eta_p)|^2}H(\eta_p)\,,
\end{multline} where 
\begin{align}
C_l(z)&=\frac{2^l e^{-\pi z/2}|\Gamma(l+1+i z)|}
{\Gamma(2l+2)}\,,\nonumber\\
H(z)&=\psi(iz)+\frac{1}{2iz}-\ln(iz)\,,
\end{align}
with $\psi(z)$ the digamma function. The ERE is reproduced in halo EFT with contact interactions embedded in the Coulomb field. In the supplementary material~\cite{supp}, we relate the EFT couplings to the ERE parameters. 

The Coulomb-subtracted $d$-wave phase shift is known to be small~\cite{Boykin:1972} and as such only the Coulomb forces are used to describe initial state $d$-wave scattering. Naively, $d$-wave operators involve higher derivatives, and thus, they are suppressed at low energy. $d$-wave contributions without short-ranged strong interaction is included at NLO~\cite{Higa:2016igc, Premarathna:2019tup} and strong interaction contributions are expected to contribute at beyond NNLO  as the corresponding operators involve more than one derivative. 

The initial $s$-wave scattering state is proportional to
\begin{multline}
\frac{1}{ [C_0(\eta_p)]^2 p(\cot\delta_0-i )}\approx
\left[-\frac{1}{a_0} +\frac{r_0}{2}p^2 -2\kappa_C H(\eta_p)\right]^{-1}\\
-\frac{s_0}{4}p^4\left[-\frac{1}{a_0} +\frac{r_0}{2}p^2 -2\kappa_C H(\eta_p)\right]^{-2}\\+
\frac{s_0^2}{16}p^8\left[-\frac{1}{a_0} +\frac{r_0}{2}p^2 -2\kappa_C H(\eta_p)\right]^{-3}+\dots\,,
\end{multline}
where we used the EFT power counting~\cite{Higa:2016igc,Premarathna:2019tup} estimate $1/a_0\sim Q^3/\Lambda^2\sim r_0/2-2\kappa_C H(\eta_p)$ and $s_0\sim \Lambda^3$ for an expansion in $Q/\Lambda$. The important observation is that a NNLO calculation does not involve any new $s$-wave scattering parameters beyond 
 $a_0$, $r_0$ and $s_0$ which already appear at NLO.

The E1 transition operator $\operatorname{O}_\text{E1}$ is comprised of both one-body and two-body currents. The former is obtained by gauging the momentum $\bm{p}\rightarrow \bm{p}+Z e \bm{A}$ where $Z$ is the charge of the particle in units of the proton charge $e$. The one-body current has no unknown coupling. Usually, two-body currents are suppressed. However, in this system fine tuning in the $s$-wave scattering length $a_0$  in the presence of the Coulomb interaction enhances a set of two-body currents to leading order (LO)~\cite{Higa:2016igc,Premarathna:2019tup}. Higher-order corrections to the two-body currents would involve two-derivatives, making these NNLO contributions. 

The LO currents were written in Ref.~\cite{Higa:2016igc} as
\begin{multline}\label{eq:LOcurrent}
e\mu\left( \frac{ Z_\phi}{m_\phi}-\frac{ Z_\psi}{m_\psi}\right)\sqrt{\frac{6\pi}{\mu}}\sqrt{\frac{2\pi}{\mu}}
L^{(\zeta)}_\text{E1} \\
\times\sum_{\alpha,\beta=1}^2\sum_{i,j=1}^3
{\chi^{\alpha,\zeta}_i}^\dagger P_{ij}^{\alpha\beta,\zeta}\chi^{\beta,s} E_j +\operatorname{h.c.}\,,
\end{multline}
where $\bm{E}$ is the electric field, $\chi^{\alpha,s}$ is an auxiliary/dimer field with spin index $\alpha$ for the initial $^2S_{1/2}$ state; 
$\chi^{\alpha,\zeta}_i$ is the dimer field with spin index $\alpha$ and vector index $i$  for the final $p$-wave bound state;  and $L^{(\zeta)}_\text{E1}$ is the two-body current coupling. A gaussian integration over the $s$-wave dimer field results in an operator proportional to 
\begin{multline}
 L^{(\zeta)}_\text{E1}  \sum_{\alpha,\beta=1}^2\sum_{i,j=1}^3 \left[\psi^\alpha\P_i \!\phi\right]^\dagger P_{ij}^{\alpha\beta,\zeta}
    \psi^\beta \phi E_j\\ +\operatorname{h.c.}
\end{multline}
Renormalization of the two-body current coupling $L^{(\zeta)}_\text{E1}$, which is treated non-perturbatively, is simpler in the dimer language and thus we continue to work in that formalism. However, looking at the interaction without the dimer is more intuitive.   

At NNLO, there are two new two-body currents in each of the final $p$-wave channels $\zeta=0$, 1 that we write as
\begin{multline}\label{eq:NNLOcurrent}
2e \mu^2\left( \frac{Z_\phi}{m_\phi}-\frac{Z_\psi}{m_\psi}\right)\sqrt{\frac{6\pi}{\mu}}\sqrt{\frac{2\pi}{\mu}}
L^{(\zeta,s)}_\mathrm{E1}\\ \times \!
\sum_{\alpha,\beta=1}^2\sum_{i,j=1}^3
{\chi^{\alpha,\zeta}_i}^\dagger P_{ij}^{\alpha\beta,\zeta}E_j \left(i\partial_0+\frac{\nabla^2}{2M}\right)\chi^{\beta,s}  \\
+ 2e \mu^2\left( \frac{Z_\phi}{m_\phi}-\frac{Z_\psi}{m_\psi}\right)\sqrt{\frac{6\pi}{\mu}}\sqrt{\frac{2\pi}{\mu}}
L^{(\zeta,p)}_\mathrm{E1} \sum_{\alpha,\beta=1}^2 \sum_{i,j=1}^3 \\ \times 
\left[\left(i\partial_0+\frac{\nabla^2}{2M}\right)\chi^{\alpha,\zeta}_i\right]^\dagger P_{ij}^{\alpha\beta,\zeta}E_j \chi^{\beta,s}  +\operatorname{h.c.}\,,
\end{multline}
where the notation for the two-body current couplings $L^{(\zeta,s)}_\mathrm{E1}$ and $L^{(\zeta,p)}_\mathrm{E1}$ become clear when we integrate the dimer fields out of the theory. We find the new operators proportional to  
\begin{multline}
L^{(\zeta,s)}_\mathrm{E1}\sum_{\alpha,\beta=1}^2\sum_{i,j=1}^3
    \left[\psi^{(\alpha)}\P_i\phi\right]^\dagger \\ \times P_{ij}^{\alpha\beta,\zeta}
\left[\psi^{(\beta)} \P^2\phi \right]E_j+\operatorname{h.c.}\,,\hspace{0.1in}\text{and}\\
L^{(\zeta,p)}_\mathrm{E1} 
\sum_{\alpha,\beta=1}^2\sum_{i,j=1}^3
 \left[\psi^{(\alpha)}\P_i \right. \\ \times \left. \P^2\phi \right]^\dagger P_{ij}^{\alpha\beta,\zeta}
\left[\psi^{(\beta)} \phi \right]E_j
+\operatorname{h.c.}\,,
\end{multline}
respectively. 
The NNLO operators with couplings $L^{(\zeta,s)}_\mathrm{E1}$, $L^{(\zeta,p)}_\mathrm{E1}$  have two extra derivatives on the $s$- and $p$-wave configurations, respectively, compared to the LO operator in Eq.~\ref{eq:LOcurrent}.
Higher order two-body currents involve higher powers of $i\partial_0+\nabla^2/(2M)$ on the dimer propagators.

The new NNLO contributions to the cross section come from the two-body currents in Eq.~\ref{eq:NNLOcurrent}. These currents satisfy gauge and Galilean invariance which become apparent after integrating out the dimer fields as above. The renormalization conditions for the two-body currents in the dimer language are trivial: $L^{(\zeta)}_\mathrm{E1}\!\sim\! 1$, $L^{(\zeta,s)}_\mathrm{E1}\!\sim\!1/\Lambda^2\!\sim\! L^{(\zeta,p)}_\mathrm{E1}$.

The final bound state $|\psi_\text{bound}\rangle$ contributes towards the wave function renormalization constant. It is parameterized in EFT 
in terms of the binding momentum $\gamma_\zeta$ and the $p$-wave effective momentum $\rho_1^{(\zeta)}\sim Q$ in the ERE~\cite{Higa:2016igc}, and the form remains unchanged at NNLO. Thus, no new calculation is need and we write~\cite{Higa:2016igc}
\begin{multline}\label{eq:wavefunction}
    \frac{1}{\mathcal Z^{(\zeta)}}
    =-\frac{1}{\mu}\left\{\rho_1^{(\zeta)}-4k_C\,H\left(-i\frac{k_C}{\gamma_\zeta}\right)\right.\\
    \left.
-\frac{2k_C^2}{\gamma^3}(k_C^2-\gamma^2)
\left[\psi'\left(\frac{k_C}{\gamma}\right)
-\frac{\gamma^2}{2k_C^2}-\frac{\gamma}{k_C}\right]\right\}\,.
\end{multline}
In the supplementary material~\cite{supp} we provide a derivation of $\mathcal Z^{(\zeta)}$.

The capture from $d$-wave initial state remains unchanged at NNLO. We have from Ref.~\cite{Higa:2016igc}  
\begin{multline}\label{eq:Dwavecapture}
    D^{(\zeta)}(p)= -\mu e^{i\sigma_2(p)}\frac{2\gamma_\zeta}{3}\Gamma(2+\frac{k_C}{\gamma_\zeta})\\ \times\int_0^\infty dr\,  r\, W_{-\frac{k_C}{\gamma_\zeta},\frac{3}{2}}(2 \gamma_\zeta r)
    \left(\frac{\partial}{\partial r} +\frac{3}{r}\right) \frac{F_2(\eta_p, r p)}{r p}\,. 
\end{multline}
$F_l(\eta_p, rp)$ is the regular Coulomb wave function~\cite{nist} and $W_{k,\mu}(z)$ is the conventionally defined Whittaker function. The Coulomb phase shifts are given by $\sigma_l(p)=\operatorname{arg}\Gamma(l+1+i\eta_p)$. 

The $s$-wave contribution is~\cite{Higa:2016igc}
\begin{multline}\label{eq:Swavecapture}
S^{(\zeta)}(p)=
-e^{i\sigma_0} C_0(\eta_p)\,
\Bigg[X(p)\\
-\frac{2\pi}{\mu^2}
\frac{B(p)+ \mu J_0(p) }
{[C_0(\eta_p)]^2 (p\cot\delta_0-ip )}\\
\left.
-2\pi k_0\frac{L^{(\zeta)}_\mathrm{E1} +p^2 L^{(\zeta,s)}_\mathrm{E1} 
    -\gamma_\zeta^2 L^{(\zeta,p)}_\mathrm{E1} }
{[C_0(\eta_p)]^2 (p\cot\delta_0-ip )}
\right]\,,
\end{multline}
where $k_0=(p^2+\gamma_\zeta^2)/(2\mu)$ is the photon energy. We estimate $\mu\sim \Lambda^2/Q$ and thus $k_0\sim Q^3/\Lambda^2$. 

The first term in Eq.~(\ref{eq:Swavecapture}) is the contribution from the purely Coulomb interaction in the initial state
\begin{multline}
  X(p)=1+
\frac{2\gamma}{3}\frac{\Gamma(2+\frac{k_C}{\gamma_\zeta})}{C_0(\eta_p)}\\
\times\int_0^\infty dr\ r 
W_{-\frac{k_C}{\gamma_\zeta},\frac{3}{2}}(2\gamma_\zeta r)\partial_r
\left[ \frac{F_0(\eta_p, r p)}{r p} \right]\,,
\end{multline}
which contributes at LO. 
The second term proportional to integrals $B(p)$ and $J_0(p)$ describes contributions from initial state strong and Coulomb interactions~\cite{Higa:2016igc}, and contributes at LO. The linear combination~\cite{Higa:2016igc,Premarathna:2019tup}
\begin{multline}\label{eq:BmuJ}
B(p)+\mu J_0(p) =\frac{\mu^2}{3\pi}\frac{ip^3-\gamma_\zeta^3}{p^2+\gamma_\zeta^2}+k_C C(p)+\Delta B(p)\\
-\frac{k_C\mu^2}{2\pi}\left[2 H(\eta_p)+2\gamma_E -\frac{5}{3}+ \ln 4\pi
\right]\,,
\end{multline}
remains finite, in which
\begin{multline}
     C(p)=\frac{\mu^2}{6\pi^2(p^2+\gamma_\zeta^2)}\int_0^1dx\int_0^1dy
  \frac{1}{\sqrt{x(1-x)}\sqrt{1-y}}\\
  \times \left[ xp^2
    f(p^2,-\gamma_\zeta^2;x,y)
  +p^2 
    f(p^2,p^2;x,y) \right.\\
  \left. +x\gamma_\zeta^2f(-\gamma_\zeta^2,-\gamma_\zeta^2;x,y)  +\gamma_\zeta^2f(-\gamma_\zeta^2,p^2;x,y)
  \right]\, ,\\
  f(a,b;x,y)=\ln \left\{\frac{\pi}{4 k_C^2}[-ya -(1-y) \frac{b}{x} -i0^+]\right\}\,.
\end{multline}
$\Delta B(p)$ is obtained from the integral~\cite{Higa:2016igc} 
\begin{multline}
  B(p)=-\frac{\mu^2\gamma_\zeta}{12\pi^2}\Gamma(2\!+\!\frac{k_C}{\gamma_\zeta})\Gamma(1+i\eta_p)\\
  \times\int_0^\infty dr\, r\,
  W_{\!-\frac{k_C}{\gamma_\zeta},\frac{3}{2}}(2\gamma_\zeta r)\frac{\partial}{\partial r}\frac{W_{-i\eta_p,\frac{1}{2}}(-i2 p r)}{r}\,,
\end{multline}
by subtracting the zero and single Coulomb photon-exchange diagrams that are calculated in Eq.~(\ref{eq:BmuJ}) explicitly. 
The origin-to-origin Coulomb Green's function $J_0(p)=G_C^{(+)}(E;0,0)$ is derived in the supplementary material~\cite{supp}.

\begin{figure}[htb]
\begin{center}
\includegraphics[width=0.25\textwidth,clip=true]{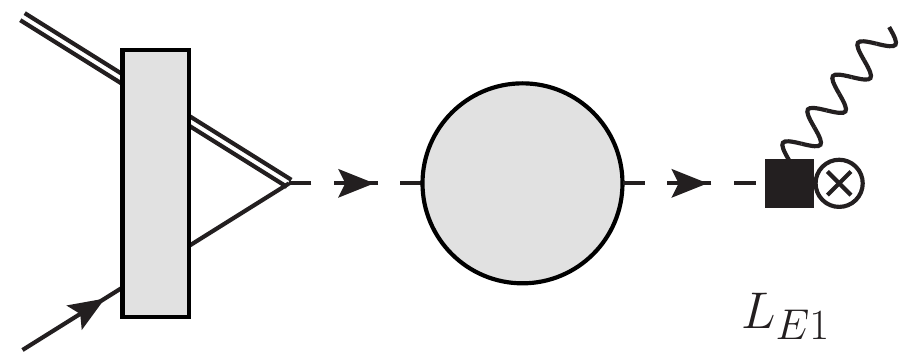}
\end{center}
\caption{\protect Two-body currents. The double line represents the scalar $\alpha$, the single line with an arrow the fermionic $^3$He, the wavy line the outgoing photon, and the dashed lines with an arrow the  dimer field in $s$-wave. The gray rectangle represents Coulomb interaction and the gray blob the strong and Coulomb interactions. The $\otimes$ represents the final $p$-wave bound state and $\blacksquare$ the two-body currents at LO and NNLO as appropriate.}
\label{fig:twobody}
\end{figure}

The third term in Eq.~(\ref{eq:Swavecapture}) constitutes two-body current contributions from the diagram in  Fig.~\ref{fig:twobody}. The NNLO contributions are very similar to the LO contributions calculated earlier~\cite{Higa:2016igc}, except that they now involve two extra derivatives. The operator with coupling $L_\text{E1}^{(\zeta,s)}$ involves derivatives on the incoming $s$-wave state with relative cm momentum $p$, hence, they involve a factor of $p^2$ relative to the LO contribution with two-body current coupling $L_\text{E1}^{(\zeta)}$. Similarly, the operator with coupling 
$L_\text{E1}^{(\zeta,p)}$ involves derivatives on the final  $p$-wave bound state with binding momentum $\gamma_\zeta$, hence, they involve a factor of $-\gamma_\zeta^2$ relative to the LO contribution.

The expressions in Eqs.~(\ref{eq:Dwavecapture}) and (\ref{eq:Swavecapture}) are used to calculate the S-factor $S_{34}(p)$ and branching ratio $R(p)$ in Eqs~(\ref{eq:S34}) and (\ref{eq:BR}), respectively. The EFT expressions for capture depend on unknown $s$-wave scattering parameters $a_0$, $r_0$, $s_0$; $p$-wave scattering parameters $\rho_1^{(\zeta)}$, and two-body current couplings $L_\text{E1}^{(\zeta)}$, $L_\text{E1}^{(s, \zeta)}$ and $L_\text{E1}^{(p,\zeta)}$. We determine the couplings and parameters from fits to capture data in the next section.  The S-factor  and branching ratio are expanded perturbatively in $Q/\Lambda$ to NNLO in the fits. We also fit the scattering parameters to available phase shift data by Boykin et al.~\cite{Boykin:1972} which is  a reanalysis of an older data set~\cite{Spiger:1967}. The phase shifts are within the EFT range of applicability, though, available only at higher energies. Thus,  we fit to phase shift without a $Q/\Lambda$ expansion. Previous analysis in Refs.~\cite{Higa:2016igc,Premarathna:2019tup} show that fits with and without the phase shift data give comparable results.  Since phase shifts add more experimental information to the analysis, we keep them.

\section{Results and Analysis}
\label{sec:results}

To make a meaningful comparison with the NLO result reported in SF3~\cite{Acharya:2024lke}, we fit the EFT parameters at NNLO to the same data set using the same fitting procedure. The parameters are determined as posterior distributions $P(\bm\theta|D)$ in a Bayesian analysis where $\bm\theta$ indicates the set of fitting parameters given the data $D$.  From Bayes' theorem, the posterior is related to the likelihood function $P(D|\bm\theta)$, written as a conditional probability for the data $D$ given theory parameters $\bm\theta$ as 
$P(\bm\theta|D)=P(D|\bm\theta) P(\bm\theta)/P(D)$, where $P(\bm\theta)$ is the prior distributions of the fit parameters.  
The likelihood function $P(D|\bm\theta)\propto \exp(-\chi^2/2)$ with the $\chi^2$ expressed as~\cite{DAgostini:1994}
\begin{align}\label{eq:chisq}
\chi^2 =\sum_{\alpha=1}^{N_\text{set}}\left\{\sum_{i=1}^{N_\alpha}{ \displaystyle
\frac{[y_{\alpha,i}\!-\!{\mu_{\alpha,i}(\bm\theta)}/{s_\alpha}]^2}{\sigma_{\alpha,i}^2}}+
\frac{(1\!-\!s_\alpha)^2}{\omega_\alpha^2}\right\}\,,
\end{align}
whereat $\alpha=1,\dots, N_{\text{set}}$ labels the different  data sets and $i=1,\dots,N_\alpha$ the individual data points in each set. The common-mode-error (CME) $\omega_\alpha$ for each data set represents systematic and other errors that are nearly energy independent.  
The CMEs are used to address the relative normalization $s_\alpha$ between the data sets that are found to describe the same energy dependence but differing in the absolute cross section. Thus along with the EFT fit parameters, we fit the normalizations with a prior  
$ s_\alpha\sim\mathcal N(1,\omega_\alpha)$  drawn from a normal distribution. The CMEs $\omega_\alpha$ are estimated in SF3~\cite{Acharya:2024lke} for S-factor activation data---ATOMKI~\cite{Bordeanu:2013}, ERNA~\cite{diLeva:2009}, LUNA~\cite{Bemmerer:2006,Confortola:2007,Gyurky:2007}, Madrid~\cite{Carmona:2012}, Seattle~\cite{Brown:2007}, and Weizmann~\cite{Singh:2004}, see Table~\ref{table:normsS34Bayesian}. 

Branching ratio data is taken from prompt photon measurements---ERNA~\cite{diLeva:2009}, LUNA~\cite{Confortola:2007}, Notre Dame~\cite{Kontos:2013} and Seattle~\cite{Brown:2007}. Scaling factors are not needed for these since they involve ratios of cross sections. 

The priors for the EFT parameters are detailed in the supplementary material~\cite{supp}. These were chosen to be consistent with the EFT power counting~\cite{Higa:2016igc,Premarathna:2019tup,Acharya:2024lke} discussed in section~\ref{sec:perturbation}. 
The $p$-wave shape-like parameters $\sigma_1^{(\zeta)}\sim1/\Lambda$ do not contribute to the capture but they appear in the $p$-wave phase shift. The systematic error 
in the phase shift data is not well documented~\cite{Spiger:1967,Boykin:1972}.  We accommodate unknown errors in the phase shift data by inflating the reported uncertainty as $\sigma^2\rightarrow \sigma^2+K^2$ with $K\sim\mathcal U(0^\circ,10^\circ)$ drawn from a uniform distribution~\cite{Higa:2016igc,Premarathna:2019tup} consistent with the reported fitting uncertainty~\cite{Boykin:1972}.

\begin{table}[htb]
\centering
\caption{\protect The median along with the difference from the 16th and 84th percentile of EFT parameter fits. The ANCs $C_{1,\zeta}^2$ are derived from the $p$-wave effective momenta $\rho_1^{(\zeta)}$.}
\begin{ruledtabular}
\begin{tabular}{llll}
Parameters
& LO &  NLO & NNLO 
\\ \hline \rule{0pt}{0.9\normalbaselineskip}
\begin{filecontents*}{params-Transpose-Bayesian-2000keV-2025-8-31.csv}
Params,LO,dLOup,dLOdown,NLO,dNLOup,dNLOdown,NNLO,dNNLOup,dNNLOdown
$a_0\, (\mathrm{fm})$,41,6,8,30,5,2,42,8,7
$r_0\, (\mathrm{fm})$,0.89,0.11,0.11,1.05,0.09,0.09,0.98,0.09,0.08
$s_0\, (\mathrm{fm}^3)$,9999,5555,5555,-1.2,0.9,0.8,-1.3,0.8,0.8
$\rho_1^{(0)}\, (\mathrm{MeV})$,-59,2,5,-55,1,2,-58,3,3
$C_{1,0}^2\, (\mathrm{fm}^{-1})$,9.5,2.5,2.6,15.2,2.9,3.2,11.3,3.4,2.9
$\sigma_1^{(0)}\, (\mathrm{fm})$,9999,5555,5555,1.57,0.07,0.05,1.65,0.11,0.08
$\rho_1^{(1)}\, (\mathrm{MeV})$,-44,2,2,-41,1,2,-40,1,2
$C_{1,1}^2\, (\mathrm{fm}^{-1})$,8.7,1.3,1.4,11.7,1.7,2,13,3.2,2.5
$\sigma_1^{(1)}\, (\mathrm{fm})$,9999,5555,5555,1.71,0.07,0.06,1.68,0.07,0.07
$L_\text{E1}^{(0)}$,1.06,0.16,0.12,0.79,0.13,0.08,1,0.13,0.13
$L_\text{E1}^{(1)}$,1.04,0.11,0.08,0.78,0.13,0.09,0.8,0.12,0.12
$L_\text{E1}^{(0,s)}\, (\mathrm{fm}^2)$,9999,5555,5555,9999,5555,5555,-0.17,0.18,0.15
$L_\text{E1}^{(0,p)}\, (\mathrm{fm}^2)$,9999,5555,5555,9999,5555,5555,0.29,0.4,0.42
$L_\text{E1}^{(1,s)}\, (\mathrm{fm}^2)$,9999,5555,5555,9999,5555,5555,-0.06,0.22,0.21
$L_\text{E1}^{(1,p)}\, (\mathrm{fm}^2)$,9999,5555,5555,9999,5555,5555,0.05,0.45,0.45
\end{filecontents*}
\csvreader[head to column names, late after line=\\]{params-Transpose-Bayesian-2000keV-2025-8-31.csv}{}
{\ \Params
& 
\ifthenelse{\equal{\LO}{9999}}{---}
{$\LO^{+\dLOup}_{-\dLOdown}$}
& 
\ifthenelse{\equal{\NLO}{9999}}{---}
{$\NLO^{+\dNLOup}_{-\dNLOdown}$}
& 
\ifthenelse{\equal{\NNLO}{9999}}{---}
{$\NNLO^{+\dNNLOup}_{-\dNNLOdown}$}
} 
\end{tabular}
\end{ruledtabular}
 \label{table:paramsS34Bayesian}
\end{table}
The fitted value of the EFT parameters are shown in Table~\ref{table:paramsS34Bayesian}. The LO expression was fitted to capture data with $E\leq \SI{1000}{\kilo\eV}$ and does not include any $d$-wave contribution that enter at NLO. The NLO and NNLO expressions were fitted to capture data with $E\leq\SI{2000}{\kilo\eV}$. We use phase shift data with $E\leq \SI{3000}{\kilo\eV}$. We see that the fitted EFT parameters are consistent with the power counting with changes from order to order within the $Q/\Lambda\sim 1/3$. In the fits, the LO and NLO values are from those reported earlier in SF3~\cite{Acharya:2024lke}. The NNLO results are new. We also find that the scaling factors for the data consistent with the reported CMEs, Table~\ref{table:normsS34Bayesian}. These consistency checks are necessary to build confidence in the theoretical error estimates. 
\begin{table}[htb]
\centering
\caption{Mean and standard deviation of the posterior of the scaling factors for the various data sets from Bayesian fits. For Boykin data, inflated errors $K$ instead of a scaling factor.}
\begin{ruledtabular}
\begin{tabular}{lllll}
Data & CME
& LO  &  NLO & NNLO 
\\ \hline
\begin{filecontents*}{norms-Transpose-Bayesian-2000keV-2025-8-31.csv}
Data,CME,LO,dLO,NLO,dNLO,NNLO,dNNLO
ATOMKI,5.9,1,5,1.01,3,1.01,3
ERNA,5.0,0.96,3,0.95,2,0.95,2
LUNA,3.0,1.02,2,1.02,2,1.02,2
Madrid,5.2,1,3,0.99,3,0.99,3
Seattle,3.0,0.96,2,0.96,1,0.95,1
Weizmann,2.2,1.02,2,1.03,2,1.02,2
Boykin (deg),9999,9.3,5,0.28,21,0.29,22
\end{filecontents*}
\csvreader[head to column names, late after line=\\]{norms-Transpose-Bayesian-2000keV-2025-8-31.csv}{}
{\ \Data
& 
\ifthenelse{\equal{\CME}{9999}}{---}{\CME\%}
& 
\ifthenelse{\equal{\LO}{9999}}{---}
{$\LO(\dLO)$}
& 
\ifthenelse{\equal{\NLO}{9999}}{---}
{$\NLO(\dNLO)$}
& 
\ifthenelse{\equal{\NNLO}{9999}}{---}
{$\NNLO(\dNNLO)$}
} 
\end{tabular}
\end{ruledtabular}
 \label{table:normsS34Bayesian}
\end{table}  

In Table~\ref{table:normsS34Bayesian}, we also provide values for ANCs
\begin{align}
C_{1,\zeta}^2= \frac{\gamma_\zeta^2[\Gamma(2+k_C/\gamma_\zeta)]^2}{\pi}\frac{2\pi}{\mu} \mathcal Z^{(\zeta)}\,,
\end{align}
which are proportional to the wave function renormalization constants $\mathcal Z^{(\zeta)}$~\cite{Higa:2016igc,Premarathna:2019tup,Higa:2020vmm}, and thus depend on the effective momenta $\rho_1^{(\zeta)}$.

\begin{figure}[htb]
\begin{center}
\includegraphics[width=\columnwidth,clip=true]{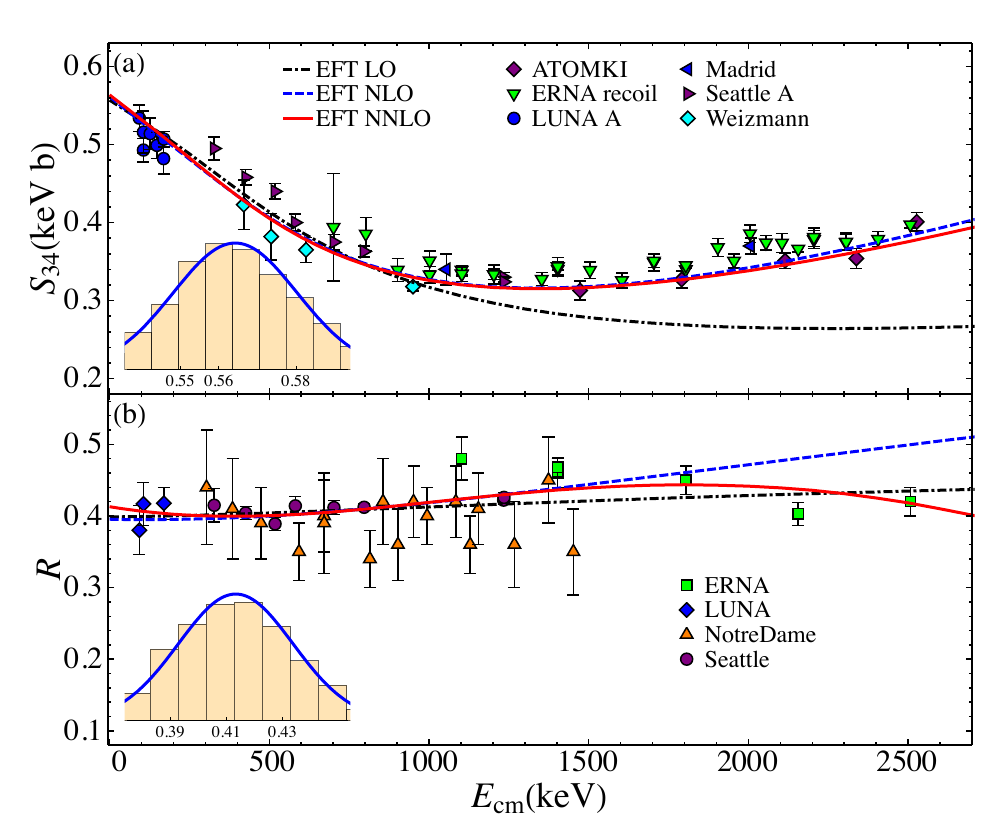}
\end{center}
\caption{\protect S-factor and branching ratio $R$ from Bayesian fits. The curves show the median values.  The insets are the  NNLO posterior distributions for $S_{34}$ and $R$ at $E=60\times 10^{-3}$ keV.  
}
\label{fig:S34Bayesian}
\end{figure}

The fit to capture data is shown in Fig.~\ref{fig:S34Bayesian}. The S-factor $S_{34}(p)$ at NNLO is nearly identical to the NLO result. This is not surprising since the NLO result already describe the capture data in the fitted region $E\leq\SI{2000}{\kilo\eV}$ accurately. There are no unique features that NNLO has to describe that is missed by the NLO expression. We find that the branching ratio $R(p)$ at NNLO is qualitatively different. It was observed earlier in Ref.~\cite{Higa:2016igc} that in the total cross section, one can compensate the effect of the two-body currents by the wave function renormalization factor to a large extend. However, the branching ratio was found to be more sensitive to the momentum dependence of the two-body-current contribution. This is reflected in the NNLO result where the two-body current couplings $L_\text{E1}^{(p,\zeta)}$ bring in extra momentum dependence. Nevertheless, one should note that the highest two ERNA data points for $R(p)$ are not used in the fits since they are above $\SI{2000}{\kilo\eV}$ and both NLO and NNLO are consistent with the fitted branching ratio data. An attractive feature of Bayesian analysis is that fits with extra parameters involve a penalty in the posterior distribution calculation, providing a balance between including enough parameters to capture the features in the data and no more~\cite{Sivia}. Indeed, we find the relative evidences, calculated using MultiNest~\cite{Feroz:2009} implemented in Python~\cite{Nestle}, 
for the NNLO vs NLO fits to be similar $\ln[P_\text{NNLO}(D)/P_\text{NLO}(D)]=\num{-2.52\pm0.52}$, slightly favoring the NLO result. However, earlier analysis~\cite{Premarathna:2019tup,Higa:2016igc} indicate systematic variation among different implementations of Nested Sampling~\cite{Skilling:2006} for evidence calculation that are about the size of the differences in the NLO and NNLO results here.
 
\begin{figure}[htb]
\begin{center}
\includegraphics[width=\columnwidth,clip=true]{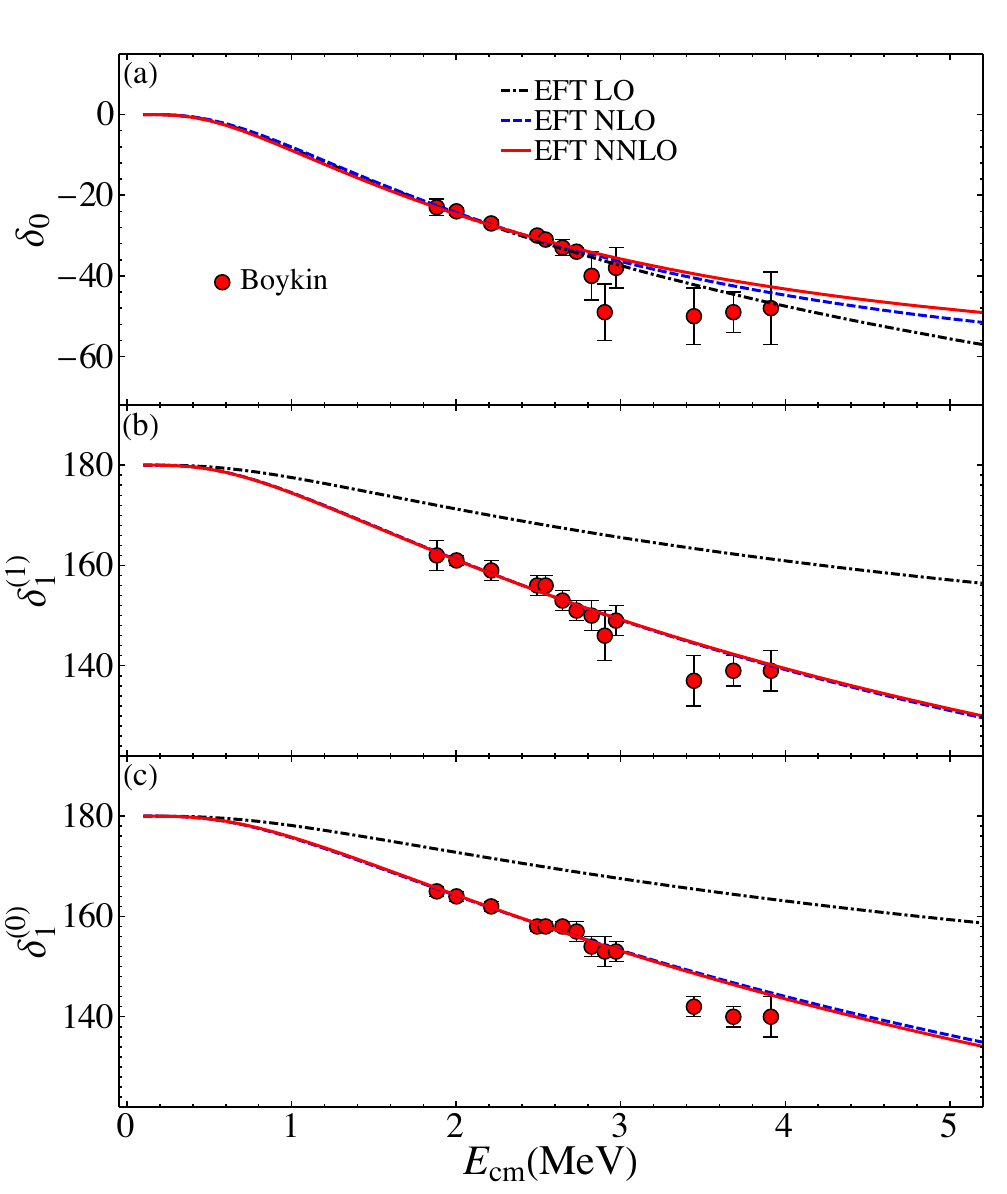}
\end{center}
\caption{\protect  Elastic scattering phase shifts (medians) from Bayesian fits for $^2S_{1/2}$, $^2P_{1/2}$, and $^2P_{3/2}$ channels, respectively, in degrees. 
}
\label{fig:PhaseShiftBayesian}
\end{figure}

The fits to the phase shift are shown in Fig.~\ref{fig:PhaseShiftBayesian}. The LO $s$-wave phase shift includes scattering length $a_0$ and effective range $r_0$, and at NLO and NNLO we include the shape parameter $s_0$. The LO $p$-wave phase shift fits involve an effective momentum parameter $\rho_1^{(\zeta)}$ in each of the channels. The higher order fits with an extra shape-like parameter $\sigma_1^{(\zeta)}$ improves the fit. Ideally, phase shift data at lower energies would be preferable. Excluding the phase shifts from the fits doesn't change the S-factor results qualitatively. 

In Table~\ref{table:S34Bayesian}, we present the S-factor and its first derivative near threshold $E_0=\SI{60e-3}{\kilo\eV}$.  The NNLO numerical values are consistent with the NLO recommendation in SF3~\cite{Acharya:2024lke}. Now, with the NNLO results, we can self-consistently estimate the theory uncertainties at around 3\%. 
\begin{table}[htb]
\centering
\caption{$S_{34}$ and its first energy derivative at $E_0=60\times10^{-3}$ keV. The first set of errors is from the Bayesian fits. The second set is the estimated LO 30\%, NLO 10\%, NNLO 3\%  EFT errors, respectively, from higher order corrections. }
\begin{ruledtabular}
\begin{tabular}{rll}
Fits & $S_{34}$ (\si{\kilo\eV\barn})
& $S'_{34}\times 10^4$ (\si{\barn})
\\  \hline \rule{0pt}{0.9\normalbaselineskip}
\begin{filecontents*}{S34-Bayesian-2000keV-2025-8-31.csv}
Theory,S,dSup,dSdown,dSEFT,Sp,dSpup,dSpdown,dSpEFT
LO,0.557,0.018,0.017,0.167,-2.6,0.6,0.6,0.8
NLO,0.561,0.017,0.018,0.056,-3.3,0.6,0.5,0.3
NNLO,0.564,0.017,0.015,0.017,-3.2,0.6,0.6,0.1
\end{filecontents*}
\csvreader[head to column names, late after line=\\]{S34-Bayesian-2000keV-2025-8-31.csv}{}
{ \Theory 
& \ifthenelse{\equal{\S}{9999}}{---}{
\ifthenelse{\equal{\dSup}{5555}}{$\S$}
{$\S^{+\dSup}_{-\dSdown}\pm\dSEFT$}}
&\ifthenelse{\equal{\Sp}{9999}}{---}{
\ifthenelse{\equal{\dSpup}{5555}}{$\Sp$}
{$\Sp^{+\dSpup}_{-\dSpdown}\pm\dSpEFT$}}
}
\end{tabular}
\end{ruledtabular}
 \label{table:S34Bayesian}
\end{table}

Finally, we provide an analytic  expression that fits the NNLO theory curve for  $E\leq \SI{1600}{\kilo \eV}$ using the same form from SF3~\cite{Acharya:2024lke} and previous work~\cite{Kajino:1987,Adelberger:2011}
\begin{multline}
     S_{34}(E)  = 
     \left[\SI{0.564\pm0.006}{\kilo\eV\barn}\right] 
     \exp\left[\num{-0.544\pm0.167} \frac{E}{\si{\kilo\eV}}\right]  \\ 
     \times    \left[ 1 - \num{0.528\pm0.464}\frac{E^2}{\si{\kilo\eV^2}} + \num{0.696\pm0.481}
     \frac{E^3}{\si{\kilo\eV^3}} \right.\\
     \left.- \num{0.176\pm0.172} \frac{E^4}{\si{\kilo\eV^4}}   \right]\,,
\end{multline}
where we indicated the uncertainty from the $\chi^2$ fit. 
S-factors at various energies used in the fit (and also branching ratios) are  in the supplementary material~\cite{supp}.

\section{Conclusions}
\label{sec:conclusions}

We perform the full NNLO Halo EFT calculation of the 
$^3\mathrm{He}(\alpha,\gamma)^7\mathrm{Be}$ reaction to aim the 
nominal 3\% theoretical uncertainties demanded by precision solar 
physics. 

Three different places are prone to receive the corresponding 
NNLO corrections ---the initial scattering states, the final $p$-wave bound 
states, and the E1 electromagnetic operator. Only in the latter new 
parameters appear. The NNLO contribution to the initial scattering state 
comes from two perturbative insertions of higher derivative operator 
proportional to the $s$-wave shape parameter $s_0$, which first appears at 
NLO. The leading $d$-wave strong interaction operator involves at least two powers 
of momentum dependence and is  subleading compared to pure 
Coulomb $d$-wave contribution which, in our power-counting, is already a NLO 
effect. This is inline with the smallness of the $d$-wave phase 
shifts~\cite{Boykin:1972}. For the final $p$-wave bound states, the same 
reasoning for initial $s$-waves apply ---the NNLO contribution comes 
from two insertions of the operator proportional to the $p$-wave shape 
parameter $\sigma_1$ that appears at NLO. Moreover, $\sigma_1$ does not 
contribute to the expressions of ANCs, but only in the $p$-wave phase 
shifts. The most relevant NNLO contribution are two higher-derivative 
two-body current couplings $L_\text{E1}^{(\zeta,s)}$, 
$L_\text{E1}^{(\zeta,p)}$ for each $p$-wave channel $\zeta=0$, 1. 

We obtain the EFT parameters from a Bayesian analysis of both 
capture data and elastic phase shifts. Fits at LO use capture data at 
$E\leq 1000$~keV while NLO and NNLO expressions were fitted to capture 
data with $E\leq 2000$~keV. We limit the fits to phase shifts to 
$E\leq 3000$~keV.
Results for elastic phase shifts (Fig.~\ref{fig:PhaseShiftBayesian}) and the astrophysical S-factor 
$S_{34}$ (upper panel of Fig.~\ref{fig:S34Bayesian}) show that, in these observables, NNLO 
corrections are quite small, indicating convergence in the EFT expansion. 
In the branching ratio $R$, NNLO corrections to NLO results 
remain small up to $E\sim 1700$~keV (bottom panel of Fig.~\ref{fig:S34Bayesian}). The small deviation close to zero 
energy is mostly due to the new two-body current term 
$\gamma_{\zeta}^2 L_\text{E1}^{(\zeta,p)}$ while at higher energies 
$E\gtrsim 1700$~keV the new term $p^2 L_\text{E1}^{(\zeta,s)}$ provides 
additional momentum dependence that goes in the right direction of 
higher-energy data. 

To conclude, we include all NNLO contributions  to 
the $^3\text{He}(\alpha,\gamma)^7\text{Be}$ capture reaction in the EFT expansion. In particular, we develop a prescription on 
how to generate higher-order two-body currents that respects gauge and 
Galilean invariance. With a conservative estimation of $\sim 1/3$ for the 
Halo EFT parameter expansion, one can state that higher-order 
contributions beyond NNLO are $\sim3\%$. This uncertainty estimate is based on the fact that the EFT curves fit data accurately with coupling  sizes that are consistent with the power counting and and the perturbative cross sections showing the expected convergence. 
This  estimate holds 
especially as one gets closer to the $^3\text{He}$-$\alpha$ threshold.

\acknowledgments
This work was partially supported by U.S.  NSF Grants PHY-2209184, PHY-2514906, 
INCT-FNA project 464898/2014-5 (RH), and FAPESP Thematic grants 2017/05660-0, 2019/07767-1, and 2020/04867-2. 


\begin{thebibliography}{39}%
\makeatletter
\providecommand \@ifxundefined [1]{%
 \@ifx{#1\undefined}
}%
\providecommand \@ifnum [1]{%
 \ifnum #1\expandafter \@firstoftwo
 \else \expandafter \@secondoftwo
 \fi
}%
\providecommand \@ifx [1]{%
 \ifx #1\expandafter \@firstoftwo
 \else \expandafter \@secondoftwo
 \fi
}%
\providecommand \natexlab [1]{#1}%
\providecommand \enquote  [1]{``#1''}%
\providecommand \bibnamefont  [1]{#1}%
\providecommand \bibfnamefont [1]{#1}%
\providecommand \citenamefont [1]{#1}%
\providecommand \href@noop [0]{\@secondoftwo}%
\providecommand \href [0]{\begingroup \@sanitize@url \@href}%
\providecommand \@href[1]{\@@startlink{#1}\@@href}%
\providecommand \@@href[1]{\endgroup#1\@@endlink}%
\providecommand \@sanitize@url [0]{\catcode `\\12\catcode `\$12\catcode
  `\&12\catcode `\#12\catcode `\^12\catcode `\_12\catcode `\%12\relax}%
\providecommand \@@startlink[1]{}%
\providecommand \@@endlink[0]{}%
\providecommand \url  [0]{\begingroup\@sanitize@url \@url }%
\providecommand \@url [1]{\endgroup\@href {#1}{\urlprefix }}%
\providecommand \urlprefix  [0]{URL }%
\providecommand \Eprint [0]{\href }%
\providecommand \doibase [0]{https://doi.org/}%
\providecommand \selectlanguage [0]{\@gobble}%
\providecommand \bibinfo  [0]{\@secondoftwo}%
\providecommand \bibfield  [0]{\@secondoftwo}%
\providecommand \translation [1]{[#1]}%
\providecommand \BibitemOpen [0]{}%
\providecommand \bibitemStop [0]{}%
\providecommand \bibitemNoStop [0]{.\EOS\space}%
\providecommand \EOS [0]{\spacefactor3000\relax}%
\providecommand \BibitemShut  [1]{\csname bibitem#1\endcsname}%
\let\auto@bib@innerbib\@empty
\bibitem [{\citenamefont {Acharya}\ \emph {et~al.}(2025)\citenamefont {Acharya}
  \emph {et~al.}}]{Acharya:2024lke}%
  \BibitemOpen
  \bibfield  {author} {\bibinfo {author} {\bibfnamefont {B.}~\bibnamefont
  {Acharya}} \emph {et~al.},\ }\bibfield  {title} {\bibinfo {title} {{Solar
  fusion III: New data and theory for hydrogen-burning stars}},\ }\href
  {https://doi.org/10.1103/8lm7-gs18} {\bibfield  {journal} {\bibinfo
  {journal} {Rev. Mod. Phys.}\ }\textbf {\bibinfo {volume} {97}},\ \bibinfo
  {pages} {035002} (\bibinfo {year} {2025})}\BibitemShut {NoStop}%
\bibitem [{\citenamefont {Bahcall}\ \emph {et~al.}(1997)\citenamefont
  {Bahcall}, \citenamefont {Pinsonneault}, \citenamefont {Basu},\ and\
  \citenamefont {Christensen-Dalsgaard}}]{PhysRevLett.78.171}%
  \BibitemOpen
  \bibfield  {author} {\bibinfo {author} {\bibfnamefont {J.~N.}\ \bibnamefont
  {Bahcall}}, \bibinfo {author} {\bibfnamefont {M.~H.}\ \bibnamefont
  {Pinsonneault}}, \bibinfo {author} {\bibfnamefont {S.}~\bibnamefont {Basu}},\
  and\ \bibinfo {author} {\bibfnamefont {J.}~\bibnamefont
  {Christensen-Dalsgaard}},\ }\bibfield  {title} {\bibinfo {title} {Are
  standard solar models reliable?},\ }\href
  {https://doi.org/10.1103/PhysRevLett.78.171} {\bibfield  {journal} {\bibinfo
  {journal} {Phys. Rev. Lett.}\ }\textbf {\bibinfo {volume} {78}},\ \bibinfo
  {pages} {171} (\bibinfo {year} {1997})}\BibitemShut {NoStop}%
\bibitem [{\citenamefont {Bahcall}\ \emph {et~al.}(2001)\citenamefont
  {Bahcall}, \citenamefont {Pinsonneault},\ and\ \citenamefont
  {Basu}}]{Bahcall:2000nu}%
  \BibitemOpen
  \bibfield  {author} {\bibinfo {author} {\bibfnamefont {J.~N.}\ \bibnamefont
  {Bahcall}}, \bibinfo {author} {\bibfnamefont {M.}~\bibnamefont
  {Pinsonneault}},\ and\ \bibinfo {author} {\bibfnamefont {S.}~\bibnamefont
  {Basu}},\ }\bibfield  {title} {\bibinfo {title} {{Solar models: Current epoch
  and time dependences, neutrinos, and helioseismological properties}},\ }\href
  {https://doi.org/10.1086/321493} {\bibfield  {journal} {\bibinfo  {journal}
  {Astrophys. J.}\ }\textbf {\bibinfo {volume} {555}},\ \bibinfo {pages} {990}
  (\bibinfo {year} {2001})}\BibitemShut {NoStop}%
\bibitem [{\citenamefont {Bahcall}\ \emph {et~al.}(2005)\citenamefont
  {Bahcall}, \citenamefont {Serenelli},\ and\ \citenamefont
  {Basu}}]{Bahcall_2005}%
  \BibitemOpen
  \bibfield  {author} {\bibinfo {author} {\bibfnamefont {J.~N.}\ \bibnamefont
  {Bahcall}}, \bibinfo {author} {\bibfnamefont {A.~M.}\ \bibnamefont
  {Serenelli}},\ and\ \bibinfo {author} {\bibfnamefont {S.}~\bibnamefont
  {Basu}},\ }\bibfield  {title} {\bibinfo {title} {New solar opacities,
  abundances, helioseismology, and neutrino fluxes},\ }\href
  {https://doi.org/10.1086/428929} {\bibfield  {journal} {\bibinfo  {journal}
  {Astrophys. J.}\ }\textbf {\bibinfo {volume} {621}},\ \bibinfo {pages} {L85}
  (\bibinfo {year} {2005})}\BibitemShut {NoStop}%
\bibitem [{\citenamefont {Fukuda}\ \emph {et~al.}(2001)\citenamefont {Fukuda}
  \emph {et~al.}}]{Fukuda:2001nj}%
  \BibitemOpen
  \bibfield  {author} {\bibinfo {author} {\bibfnamefont {S.}~\bibnamefont
  {Fukuda}} \emph {et~al.} (\bibinfo {collaboration} {Super-Kamiokande}),\
  }\bibfield  {title} {\bibinfo {title} {Solar {$^8$B} and hep neutrino
  measurements from 1258 days of {Super-Kamiokande} data},\ }\href
  {https://doi.org/10.1103/PhysRevLett.86.5651} {\bibfield  {journal} {\bibinfo
   {journal} {Phys. Rev. Lett.}\ }\textbf {\bibinfo {volume} {86}},\ \bibinfo
  {pages} {5651} (\bibinfo {year} {2001})}\BibitemShut {NoStop}%
\bibitem [{\citenamefont {Ahmad}\ \emph {et~al.}(2001)\citenamefont {Ahmad}
  \emph {et~al.}}]{Ahmad:2001an}%
  \BibitemOpen
  \bibfield  {author} {\bibinfo {author} {\bibfnamefont {Q.}~\bibnamefont
  {Ahmad}} \emph {et~al.} (\bibinfo {collaboration} {SNO}),\ }\bibfield
  {title} {\bibinfo {title} {{Measurement of the rate of $\nu_e+d\rightarrow
  p+p+e^-$ interactions produced by $^8$B solar neutrinos at the Sudbury
  Neutrino Observatory}},\ }\href
  {https://doi.org/10.1103/PhysRevLett.87.071301} {\bibfield  {journal}
  {\bibinfo  {journal} {Phys. Rev. Lett.}\ }\textbf {\bibinfo {volume} {87}},\
  \bibinfo {pages} {071301} (\bibinfo {year} {2001})}\BibitemShut {NoStop}%
\bibitem [{\citenamefont {Cyburt}\ and\ \citenamefont
  {Davids}(2008)}]{Cyburt:2008up}%
  \BibitemOpen
  \bibfield  {author} {\bibinfo {author} {\bibfnamefont {R.~H.}\ \bibnamefont
  {Cyburt}}\ and\ \bibinfo {author} {\bibfnamefont {B.}~\bibnamefont
  {Davids}},\ }\bibfield  {title} {\bibinfo {title} {Evaluation of modern
  {$^3$He$(\alpha,\gamma)^7$Be} data},\ }\href
  {https://doi.org/10.1103/PhysRevC.78.064614} {\bibfield  {journal} {\bibinfo
  {journal} {Phys. Rev.}\ }\textbf {\bibinfo {volume} {C78}},\ \bibinfo {pages}
  {064614} (\bibinfo {year} {2008})}\BibitemShut {NoStop}%
\bibitem [{\citenamefont {Bahcall}\ and\ \citenamefont
  {Ulrich}(1988)}]{Bahcall:1987jc}%
  \BibitemOpen
  \bibfield  {author} {\bibinfo {author} {\bibfnamefont {J.~N.}\ \bibnamefont
  {Bahcall}}\ and\ \bibinfo {author} {\bibfnamefont {R.~K.}\ \bibnamefont
  {Ulrich}},\ }\bibfield  {title} {\bibinfo {title} {{Solar Models, Neutrino
  Experiments and Helioseismology}},\ }\href
  {https://doi.org/10.1103/RevModPhys.60.297} {\bibfield  {journal} {\bibinfo
  {journal} {Rev. Mod. Phys.}\ }\textbf {\bibinfo {volume} {60}},\ \bibinfo
  {pages} {297} (\bibinfo {year} {1988})}\BibitemShut {NoStop}%
\bibitem [{\citenamefont {Abe}\ \emph {et~al.}(2018)\citenamefont {Abe} \emph
  {et~al.}}]{Hyper-Kamiokande:2018ofw}%
  \BibitemOpen
  \bibfield  {author} {\bibinfo {author} {\bibfnamefont {K.}~\bibnamefont
  {Abe}} \emph {et~al.} (\bibinfo {collaboration} {Hyper-Kamiokande}),\
  }\bibfield  {title} {\bibinfo {title} {{Hyper-Kamiokande Design Report}},\
  }\Eprint {https://arxiv.org/abs/1805.04163} {arXiv:1805.04163
  [physics.ins-det]}  (\bibinfo {year} {2018})\BibitemShut {NoStop}%
\bibitem [{\citenamefont {Anderson}\ \emph {et~al.}(2019)\citenamefont
  {Anderson} \emph {et~al.}}]{PhysRevD.99.012012}%
  \BibitemOpen
  \bibfield  {author} {\bibinfo {author} {\bibfnamefont {M.}~\bibnamefont
  {Anderson}} \emph {et~al.} (\bibinfo {collaboration} {$\mathrm{SNO}+$
  Collaboration}),\ }\bibfield  {title} {\bibinfo {title} {Measurement of the
  $^{8}\mathrm{B}$ solar neutrino flux in $\mathrm{SNO}+$ with very low
  backgrounds},\ }\href {https://doi.org/10.1103/PhysRevD.99.012012} {\bibfield
   {journal} {\bibinfo  {journal} {Phys. Rev. D}\ }\textbf {\bibinfo {volume}
  {99}},\ \bibinfo {pages} {012012} (\bibinfo {year} {2019})}\BibitemShut
  {NoStop}%
\bibitem [{\citenamefont {Abusleme}\ \emph {et~al.}(2021)\citenamefont
  {Abusleme} \emph {et~al.}}]{Abusleme_2021}%
  \BibitemOpen
  \bibfield  {author} {\bibinfo {author} {\bibfnamefont {A.}~\bibnamefont
  {Abusleme}} \emph {et~al.},\ }\bibfield  {title} {\bibinfo {title}
  {Feasibility and physics potential of detecting {$^8$B} solar neutrinos at
  {JUNO}},\ }\href {https://doi.org/10.1088/1674-1137/abd92a} {\bibfield
  {journal} {\bibinfo  {journal} {Chinese Physics C}\ }\textbf {\bibinfo
  {volume} {45}},\ \bibinfo {pages} {023004} (\bibinfo {year}
  {2021})}\BibitemShut {NoStop}%
\bibitem [{\citenamefont {Capozzi}\ \emph {et~al.}(2019)\citenamefont
  {Capozzi}, \citenamefont {Li}, \citenamefont {Zhu},\ and\ \citenamefont
  {Beacom}}]{PhysRevLett.123.131803}%
  \BibitemOpen
  \bibfield  {author} {\bibinfo {author} {\bibfnamefont {F.}~\bibnamefont
  {Capozzi}}, \bibinfo {author} {\bibfnamefont {S.~W.}\ \bibnamefont {Li}},
  \bibinfo {author} {\bibfnamefont {G.}~\bibnamefont {Zhu}},\ and\ \bibinfo
  {author} {\bibfnamefont {J.~F.}\ \bibnamefont {Beacom}},\ }\bibfield  {title}
  {\bibinfo {title} {{DUNE} as the next-generation solar neutrino experiment},\
  }\href {https://doi.org/10.1103/PhysRevLett.123.131803} {\bibfield  {journal}
  {\bibinfo  {journal} {Phys. Rev. Lett.}\ }\textbf {\bibinfo {volume} {123}},\
  \bibinfo {pages} {131803} (\bibinfo {year} {2019})}\BibitemShut {NoStop}%
\bibitem [{\citenamefont {Cyburt}\ \emph {et~al.}(2004)\citenamefont {Cyburt},
  \citenamefont {Fields},\ and\ \citenamefont {Olive}}]{PhysRevD.69.123519}%
  \BibitemOpen
  \bibfield  {author} {\bibinfo {author} {\bibfnamefont {R.~H.}\ \bibnamefont
  {Cyburt}}, \bibinfo {author} {\bibfnamefont {B.~D.}\ \bibnamefont {Fields}},\
  and\ \bibinfo {author} {\bibfnamefont {K.~A.}\ \bibnamefont {Olive}},\
  }\bibfield  {title} {\bibinfo {title} {Solar neutrino constraints on the
  {BBN} production of {Li}},\ }\href
  {https://doi.org/10.1103/PhysRevD.69.123519} {\bibfield  {journal} {\bibinfo
  {journal} {Phys. Rev. D}\ }\textbf {\bibinfo {volume} {69}},\ \bibinfo
  {pages} {123519} (\bibinfo {year} {2004})}\BibitemShut {NoStop}%
\bibitem [{\citenamefont {Higa}\ \emph {et~al.}(2018)\citenamefont {Higa},
  \citenamefont {Rupak},\ and\ \citenamefont {Vaghani}}]{Higa:2016igc}%
  \BibitemOpen
  \bibfield  {author} {\bibinfo {author} {\bibfnamefont {R.}~\bibnamefont
  {Higa}}, \bibinfo {author} {\bibfnamefont {G.}~\bibnamefont {Rupak}},\ and\
  \bibinfo {author} {\bibfnamefont {A.}~\bibnamefont {Vaghani}},\ }\bibfield
  {title} {\bibinfo {title} {{Radiative $^{3}$He($\alpha , \gamma$)$^{7}$Be
  reaction in halo effective field theory}},\ }\href
  {https://doi.org/10.1140/epja/i2018-12486-5} {\bibfield  {journal} {\bibinfo
  {journal} {Eur. Phys. J. A}\ }\textbf {\bibinfo {volume} {54}},\ \bibinfo
  {pages} {89} (\bibinfo {year} {2018})}\BibitemShut {NoStop}%
\bibitem [{\citenamefont {Premarathna}\ and\ \citenamefont
  {Rupak}(2020)}]{Premarathna:2019tup}%
  \BibitemOpen
  \bibfield  {author} {\bibinfo {author} {\bibfnamefont {P.}~\bibnamefont
  {Premarathna}}\ and\ \bibinfo {author} {\bibfnamefont {G.}~\bibnamefont
  {Rupak}},\ }\bibfield  {title} {\bibinfo {title} {{Bayesian analysis of
  capture reactions $^3\mathrm{He}(\alpha,\gamma)^7\mathrm{Be}$ and
  $^3\mathrm{H}(\alpha,\gamma)^7\mathrm{Li}$}},\ }\href
  {https://doi.org/10.1140/epja/s10050-020-00113-z} {\bibfield  {journal}
  {\bibinfo  {journal} {Eur. Phys. J. A}\ }\textbf {\bibinfo {volume} {56}},\
  \bibinfo {pages} {166} (\bibinfo {year} {2020})}\BibitemShut {NoStop}%
\bibitem [{\citenamefont {Zhang}\ \emph {et~al.}(2020)\citenamefont {Zhang},
  \citenamefont {Nollett},\ and\ \citenamefont {Phillips}}]{Zhang_2020}%
  \BibitemOpen
  \bibfield  {author} {\bibinfo {author} {\bibfnamefont {X.}~\bibnamefont
  {Zhang}}, \bibinfo {author} {\bibfnamefont {K.~M.}\ \bibnamefont {Nollett}},\
  and\ \bibinfo {author} {\bibfnamefont {D.~R.}\ \bibnamefont {Phillips}},\
  }\bibfield  {title} {\bibinfo {title} {{S-factor and scattering-parameter
  extractions from ${}^{3}\mathrm{He}+{}^{4}\mathrm{He}{ \rightarrow
  }^{7}\mathrm{Be}+\gamma$}},\ }\href
  {https://doi.org/10.1088/1361-6471/ab6a71} {\bibfield  {journal} {\bibinfo
  {journal} {J. Phys. G}\ }\textbf {\bibinfo {volume} {47}},\ \bibinfo {pages}
  {054002} (\bibinfo {year} {2020})}\BibitemShut {NoStop}%
\bibitem [{\citenamefont {Odell}\ \emph {et~al.}(2022)\citenamefont {Odell},
  \citenamefont {Brune}, \citenamefont {Phillips}, \citenamefont {deBoer},\
  and\ \citenamefont {Paneru}}]{Odell:2021nmp}%
  \BibitemOpen
  \bibfield  {author} {\bibinfo {author} {\bibfnamefont {D.}~\bibnamefont
  {Odell}}, \bibinfo {author} {\bibfnamefont {C.~R.}\ \bibnamefont {Brune}},
  \bibinfo {author} {\bibfnamefont {D.~R.}\ \bibnamefont {Phillips}}, \bibinfo
  {author} {\bibfnamefont {R.~J.}\ \bibnamefont {deBoer}},\ and\ \bibinfo
  {author} {\bibfnamefont {S.~N.}\ \bibnamefont {Paneru}},\ }\bibfield  {title}
  {\bibinfo {title} {Performing bayesian analyses with {AZURE2} using {BRICK}:
  An application to the 7{Be} system},\ }\href
  {https://doi.org/10.3389/fphy.2022.888476} {\bibfield  {journal} {\bibinfo
  {journal} {Front. in Phys.}\ }\textbf {\bibinfo {volume} {10}},\ \bibinfo
  {pages} {888476} (\bibinfo {year} {2022})}\BibitemShut {NoStop}%
\bibitem [{\citenamefont {Tak\'acs}\ \emph {et~al.}(2015)\citenamefont
  {Tak\'acs}, \citenamefont {Bemmerer}, \citenamefont {Sz\"ucs},\ and\
  \citenamefont {Zuber}}]{Takacs:2015}%
  \BibitemOpen
  \bibfield  {author} {\bibinfo {author} {\bibfnamefont {M.~P.}\ \bibnamefont
  {Tak\'acs}}, \bibinfo {author} {\bibfnamefont {D.}~\bibnamefont {Bemmerer}},
  \bibinfo {author} {\bibfnamefont {T.}~\bibnamefont {Sz\"ucs}},\ and\ \bibinfo
  {author} {\bibfnamefont {K.}~\bibnamefont {Zuber}},\ }\bibfield  {title}
  {\bibinfo {title} {Constraining big bang lithium production with recent solar
  neutrino data},\ }\href {https://doi.org/10.1103/PhysRevD.91.123526}
  {\bibfield  {journal} {\bibinfo  {journal} {Phys. Rev. D}\ }\textbf {\bibinfo
  {volume} {91}},\ \bibinfo {pages} {123526} (\bibinfo {year}
  {2015})}\BibitemShut {NoStop}%
\bibitem [{\citenamefont {Khadka}\ \emph {et~al.}(2025)\citenamefont {Khadka},
  \citenamefont {Gan}, \citenamefont {Higa},\ and\ \citenamefont
  {Rupak}}]{supp}%
  \BibitemOpen
  \bibfield  {author} {\bibinfo {author} {\bibfnamefont {R.}~\bibnamefont
  {Khadka}}, \bibinfo {author} {\bibfnamefont {L.}~\bibnamefont {Gan}},
  \bibinfo {author} {\bibfnamefont {R.}~\bibnamefont {Higa}},\ and\ \bibinfo
  {author} {\bibfnamefont {G.}~\bibnamefont {Rupak}},\ }\href@noop {} {}
  (\bibinfo {year} {2025}),\ \bibinfo {note} {{Supplementary material for
  \HeAlpha ~calculation at next-to-next-to-leading order.}}\BibitemShut {Stop}%
\bibitem [{\citenamefont {Boykin}\ \emph {et~al.}(1972)\citenamefont {Boykin},
  \citenamefont {Baker},\ and\ \citenamefont {Hardy}}]{Boykin:1972}%
  \BibitemOpen
  \bibfield  {author} {\bibinfo {author} {\bibfnamefont {W.~R.}\ \bibnamefont
  {Boykin}}, \bibinfo {author} {\bibfnamefont {S.~D.}\ \bibnamefont {Baker}},\
  and\ \bibinfo {author} {\bibfnamefont {D.~M.}\ \bibnamefont {Hardy}},\
  }\bibfield  {title} {\bibinfo {title} {Scattering of $^3${He} and $^4${He}
  from polarized $^3${He} between 4 and 10 {MeV}},\ }\href
  {https://doi.org/https://doi.org/10.1016/0375-9474(72)90732-4} {\bibfield
  {journal} {\bibinfo  {journal} {Nucl. Phys. A}\ }\textbf {\bibinfo {volume}
  {195}},\ \bibinfo {pages} {241} (\bibinfo {year} {1972})}\BibitemShut
  {NoStop}%
\bibitem [{\citenamefont {Thompson}(2025)}]{nist}%
  \BibitemOpen
  \bibfield  {author} {\bibinfo {author} {\bibfnamefont {I.~J.}\ \bibnamefont
  {Thompson}},\ }\href@noop {} {} (\bibinfo {year} {2025}),\ \bibinfo {note}
  {{NIST} Digital Library of Mathematical Functions, {\em ``Chapter 33: Coulomb
  Functions''}, accessed: 2025-07-28,
  \url{http://dlmf.nist.gov/33}}\BibitemShut {NoStop}%
\bibitem [{\citenamefont {Spieger}\ and\ \citenamefont
  {Tombrello}(1967)}]{Spiger:1967}%
  \BibitemOpen
  \bibfield  {author} {\bibinfo {author} {\bibfnamefont {R.~J.}\ \bibnamefont
  {Spieger}}\ and\ \bibinfo {author} {\bibfnamefont {T.~A.}\ \bibnamefont
  {Tombrello}},\ }\bibfield  {title} {\bibinfo {title} {Scattering of {He$^3$}
  by {He$^4$} and of {He$^4$} by tritium},\ }\href@noop {} {\bibfield
  {journal} {\bibinfo  {journal} {Phys. Rev.}\ }\textbf {\bibinfo {volume}
  {163}},\ \bibinfo {pages} {964} (\bibinfo {year} {1967})}\BibitemShut
  {NoStop}%
\bibitem [{\citenamefont {{D'Agostini}}(1994)}]{DAgostini:1994}%
  \BibitemOpen
  \bibfield  {author} {\bibinfo {author} {\bibfnamefont {G.}~\bibnamefont
  {{D'Agostini}}},\ }\bibfield  {title} {\bibinfo {title} {{On the use of the
  covariance matrix to fit correlated data}},\ }\href
  {https://doi.org/10.1016/0168-9002(94)90719-6} {\bibfield  {journal}
  {\bibinfo  {journal} {Nuclear Instruments and Methods in Physics Research A}\
  }\textbf {\bibinfo {volume} {346}},\ \bibinfo {pages} {306} (\bibinfo {year}
  {1994})}\BibitemShut {NoStop}%
\bibitem [{\citenamefont {Bordeanu}\ \emph {et~al.}(2013)\citenamefont
  {Bordeanu} \emph {et~al.}}]{Bordeanu:2013}%
  \BibitemOpen
  \bibfield  {author} {\bibinfo {author} {\bibfnamefont {C.}~\bibnamefont
  {Bordeanu}} \emph {et~al.},\ }\bibfield  {title} {\bibinfo {title}
  {Activation measurement of the {$^3\text{He}+^4\text{He}\rightarrow{}^7$Be}
  reaction cross section at high energies},\ }\href
  {https://doi.org/10.1016/j.nuclphysa.2013.03.012} {\bibfield  {journal}
  {\bibinfo  {journal} {Nucl. Phys. A}\ }\textbf {\bibinfo {volume} {908}},\
  \bibinfo {pages} {1} (\bibinfo {year} {2013})}\BibitemShut {NoStop}%
\bibitem [{\citenamefont {{di}~Leva}\ \emph {et~al.}(2009)\citenamefont
  {{di}~Leva} \emph {et~al.}}]{diLeva:2009}%
  \BibitemOpen
  \bibfield  {author} {\bibinfo {author} {\bibfnamefont {A.}~\bibnamefont
  {{di}~Leva}} \emph {et~al.},\ }\bibfield  {title} {\bibinfo {title} {Stellar
  and primordial nucleosynthesis of {$^7$Be}: Measurement of
  {$^3\text{He}+^4\text{He}\rightarrow{}^7$Be}},\ }\href
  {https://doi.org/10.1103/PhysRevLett.102.232502} {\bibfield  {journal}
  {\bibinfo  {journal} {Phys. Rev. Lett.}\ }\textbf {\bibinfo {volume} {102}},\
  \bibinfo {eid} {232502} (\bibinfo {year} {2009})}\BibitemShut {NoStop}%
\bibitem [{\citenamefont {Bemmerer}\ \emph {et~al.}(2006)\citenamefont
  {Bemmerer} \emph {et~al.}}]{Bemmerer:2006}%
  \BibitemOpen
  \bibfield  {author} {\bibinfo {author} {\bibfnamefont {D.}~\bibnamefont
  {Bemmerer}} \emph {et~al.},\ }\bibfield  {title} {\bibinfo {title}
  {Activation measurement of the {$^3\text{He}+^4\text{He}\rightarrow{}^7$Be}
  cross section at low energy},\ }\href
  {https://doi.org/10.1103/PhysRevLett.97.122502} {\bibfield  {journal}
  {\bibinfo  {journal} {Phys. Rev. Lett.}\ }\textbf {\bibinfo {volume} {97}},\
  \bibinfo {eid} {122502} (\bibinfo {year} {2006})}\BibitemShut {NoStop}%
\bibitem [{\citenamefont {Confortola}\ \emph {et~al.}(2007)\citenamefont
  {Confortola} \emph {et~al.}}]{Confortola:2007}%
  \BibitemOpen
  \bibfield  {author} {\bibinfo {author} {\bibfnamefont {F.}~\bibnamefont
  {Confortola}} \emph {et~al.},\ }\bibfield  {title} {\bibinfo {title}
  {Astrophysical {S} factor of the {$^3\text{He}+^4\text{He}\rightarrow{}^7$Be}
  reaction measured at low energy via detection of prompt and delayed $\gamma$
  rays},\ }\href {https://doi.org/10.1103/PhysRevC.75.065803} {\bibfield
  {journal} {\bibinfo  {journal} {Phys. Rev. Lett.}\ }\textbf {\bibinfo
  {volume} {75}},\ \bibinfo {eid} {065803} (\bibinfo {year}
  {2007})}\BibitemShut {NoStop}%
\bibitem [{\citenamefont {Gy{\"u}rky}\ \emph {et~al.}(2007)\citenamefont
  {Gy{\"u}rky} \emph {et~al.}}]{Gyurky:2007}%
  \BibitemOpen
  \bibfield  {author} {\bibinfo {author} {\bibfnamefont {G.}~\bibnamefont
  {Gy{\"u}rky}} \emph {et~al.},\ }\bibfield  {title} {\bibinfo {title}
  {{$^3\text{He}+^4\text{He}\rightarrow{}^7$Be} cross section at low
  energies},\ }\href {https://doi.org/10.1103/PhysRevC.75.035805} {\bibfield
  {journal} {\bibinfo  {journal} {Phys. Rev. C}\ }\textbf {\bibinfo {volume}
  {75}},\ \bibinfo {eid} {035805} (\bibinfo {year} {2007})}\BibitemShut
  {NoStop}%
\bibitem [{\citenamefont {Carmona-Gallardo}\ \emph {et~al.}(2012)\citenamefont
  {Carmona-Gallardo} \emph {et~al.}}]{Carmona:2012}%
  \BibitemOpen
  \bibfield  {author} {\bibinfo {author} {\bibfnamefont {M.}~\bibnamefont
  {Carmona-Gallardo}} \emph {et~al.},\ }\bibfield  {title} {\bibinfo {title}
  {New measurement of the {$^3\text{He}+^4\text{He}\rightarrow{}^7$Be} cross
  section at medium energies},\ }\href
  {https://doi.org/10.1103/PhysRevC.86.032801} {\bibfield  {journal} {\bibinfo
  {journal} {Phys. Rev. C}\ }\textbf {\bibinfo {volume} {86}},\ \bibinfo {eid}
  {032801} (\bibinfo {year} {2012})}\BibitemShut {NoStop}%
\bibitem [{\citenamefont {Brown}\ \emph {et~al.}(2007)\citenamefont {Brown}
  \emph {et~al.}}]{Brown:2007}%
  \BibitemOpen
  \bibfield  {author} {\bibinfo {author} {\bibfnamefont {T.~A.~D.}\
  \bibnamefont {Brown}} \emph {et~al.},\ }\bibfield  {title} {\bibinfo {title}
  {{$^3\text{He}+^4\text{He}\rightarrow{}^7$Be} astrophysical {S} factor},\
  }\href {https://doi.org/10.1103/PhysRevC.76.055801} {\bibfield  {journal}
  {\bibinfo  {journal} {Phys. Rev. C}\ }\textbf {\bibinfo {volume} {76}},\
  \bibinfo {eid} {055801} (\bibinfo {year} {2007})}\BibitemShut {NoStop}%
\bibitem [{\citenamefont {Singh}\ \emph {et~al.}(2004)\citenamefont {Singh},
  \citenamefont {Hass}, \citenamefont {Nir-El},\ and\ \citenamefont
  {Haquin}}]{Singh:2004}%
  \BibitemOpen
  \bibfield  {author} {\bibinfo {author} {\bibfnamefont {B.~S.}\ \bibnamefont
  {Singh}}, \bibinfo {author} {\bibfnamefont {M.}~\bibnamefont {Hass}},
  \bibinfo {author} {\bibfnamefont {Y.}~\bibnamefont {Nir-El}},\ and\ \bibinfo
  {author} {\bibfnamefont {G.}~\bibnamefont {Haquin}},\ }\bibfield  {title}
  {\bibinfo {title} {New precision measurement of the
  {$^{3}$He($^{4}$He,$\gamma){}^{7}$Be} cross section},\ }\href
  {https://doi.org/10.1103/PhysRevLett.93.262503} {\bibfield  {journal}
  {\bibinfo  {journal} {Phys. Rev. Lett.}\ }\textbf {\bibinfo {volume} {93}},\
  \bibinfo {pages} {262503} (\bibinfo {year} {2004})}\BibitemShut {NoStop}%
\bibitem [{\citenamefont {Kontos}\ \emph {et~al.}(2013)\citenamefont {Kontos}
  \emph {et~al.}}]{Kontos:2013}%
  \BibitemOpen
  \bibfield  {author} {\bibinfo {author} {\bibfnamefont {A.}~\bibnamefont
  {Kontos}} \emph {et~al.},\ }\bibfield  {title} {\bibinfo {title}
  {Astrophysical {S} factor of {$^3\mathrm{He}(\alpha,\gamma)^7\mathrm{Be}$}},\
  }\href {https://doi.org/10.1103/PhysRevC.87.065804} {\bibfield  {journal}
  {\bibinfo  {journal} {Phys. Rev. C}\ }\textbf {\bibinfo {volume} {87}},\
  \bibinfo {eid} {065804} (\bibinfo {year} {2013})}\BibitemShut {NoStop}%
\bibitem [{\citenamefont {Higa}\ \emph {et~al.}(2022)\citenamefont {Higa},
  \citenamefont {Premarathna},\ and\ \citenamefont {Rupak}}]{Higa:2020vmm}%
  \BibitemOpen
  \bibfield  {author} {\bibinfo {author} {\bibfnamefont {R.}~\bibnamefont
  {Higa}}, \bibinfo {author} {\bibfnamefont {P.}~\bibnamefont {Premarathna}},\
  and\ \bibinfo {author} {\bibfnamefont {G.}~\bibnamefont {Rupak}},\ }\bibfield
   {title} {\bibinfo {title} {Coupled-channels treatment of
  {$^7\text{Be}(p,\gamma)^8$B} in effective field theory},\ }\href
  {https://doi.org/10.1103/PhysRevC.106.014601} {\bibfield  {journal} {\bibinfo
   {journal} {Phys. Rev. C}\ }\textbf {\bibinfo {volume} {106}},\ \bibinfo
  {pages} {014601} (\bibinfo {year} {2022})}\BibitemShut {NoStop}%
\bibitem [{\citenamefont {Sivia}\ and\ \citenamefont {Skilling}(2005)}]{Sivia}%
  \BibitemOpen
  \bibfield  {author} {\bibinfo {author} {\bibfnamefont {D.~S.}\ \bibnamefont
  {Sivia}}\ and\ \bibinfo {author} {\bibfnamefont {J.}~\bibnamefont
  {Skilling}},\ }\href@noop {} {\emph {\bibinfo {title} {Data Analysis: A
  Bayesian Tutorial}}}\ (\bibinfo  {publisher} {Oxford University Press},\
  \bibinfo {address} {Oxford},\ \bibinfo {year} {2005})\BibitemShut {NoStop}%
\bibitem [{\citenamefont {{Feroz}}\ \emph {et~al.}(2009)\citenamefont
  {{Feroz}}, \citenamefont {{Hobson}},\ and\ \citenamefont
  {{Bridges}}}]{Feroz:2009}%
  \BibitemOpen
  \bibfield  {author} {\bibinfo {author} {\bibfnamefont {F.}~\bibnamefont
  {{Feroz}}}, \bibinfo {author} {\bibfnamefont {M.~P.}\ \bibnamefont
  {{Hobson}}},\ and\ \bibinfo {author} {\bibfnamefont {M.}~\bibnamefont
  {{Bridges}}},\ }\bibfield  {title} {\bibinfo {title} {{MULTINEST:} an
  efficient and robust bayesian inference tool for cosmology and particle
  physics},\ }\href {https://doi.org/10.1111/j.1365-2966.2009.14548.x}
  {\bibfield  {journal} {\bibinfo  {journal} {Mon. Not. R. Astron. Soc.}\
  }\textbf {\bibinfo {volume} {398}},\ \bibinfo {pages} {1601} (\bibinfo {year}
  {2009})}\BibitemShut {NoStop}%
\bibitem [{\citenamefont {Barbary}(2019)}]{Nestle}%
  \BibitemOpen
  \bibfield  {author} {\bibinfo {author} {\bibfnamefont {K.}~\bibnamefont
  {Barbary}},\ }\href@noop {} {}\bibinfo {howpublished}
  {\url{https://github.com/kbarbary/nestle}} (\bibinfo {year} {2019}),\
  \bibinfo {note} {nestle}\BibitemShut {NoStop}%
\bibitem [{\citenamefont {Skilling}(2006)}]{Skilling:2006}%
  \BibitemOpen
  \bibfield  {author} {\bibinfo {author} {\bibfnamefont {J.}~\bibnamefont
  {Skilling}},\ }\bibfield  {title} {\bibinfo {title} {Nested sampling for
  general {Bayesian} computation},\ }\href {https://doi.org/10.1214/06-BA127}
  {\bibfield  {journal} {\bibinfo  {journal} {Bayesian Analysis}\ }\textbf
  {\bibinfo {volume} {1}},\ \bibinfo {pages} {833} (\bibinfo {year}
  {2006})}\BibitemShut {NoStop}%
\bibitem [{\citenamefont {Kajino}\ \emph {et~al.}(1987)\citenamefont {Kajino},
  \citenamefont {Toki},\ and\ \citenamefont {Austin}}]{Kajino:1987}%
  \BibitemOpen
  \bibfield  {author} {\bibinfo {author} {\bibfnamefont {T.}~\bibnamefont
  {Kajino}}, \bibinfo {author} {\bibfnamefont {H.}~\bibnamefont {Toki}},\ and\
  \bibinfo {author} {\bibfnamefont {S.~M.}\ \bibnamefont {Austin}},\ }\bibfield
   {title} {\bibinfo {title} {Radiative alpha-capture rates leading to {$A=7$}
  nuclei: Applications to the solar neutrino problem and big bang
  nucleosynthesis},\ }\href {https://doi.org/10.1086/165476} {\bibfield
  {journal} {\bibinfo  {journal} {Astrophys. J.}\ }\textbf {\bibinfo {volume}
  {319}},\ \bibinfo {pages} {531} (\bibinfo {year} {1987})}\BibitemShut
  {NoStop}%
\bibitem [{\citenamefont {Adelberger}\ \emph {et~al.}(2011)\citenamefont
  {Adelberger} \emph {et~al.}}]{Adelberger:2011}%
  \BibitemOpen
  \bibfield  {author} {\bibinfo {author} {\bibfnamefont {E.~G.}\ \bibnamefont
  {Adelberger}} \emph {et~al.},\ }\bibfield  {title} {\bibinfo {title} {Solar
  fusion cross sections. {II.} {The} $pp$ chain and {CNO} cycles},\ }\href
  {https://doi.org/10.1103/RevModPhys.83.195} {\bibfield  {journal} {\bibinfo
  {journal} {Rev. Mod. Phys.}\ }\textbf {\bibinfo {volume} {83}},\ \bibinfo
  {pages} {195} (\bibinfo {year} {2011})}\BibitemShut {NoStop}%
\end{thebibliography}
%

\end{document}


\title{\texorpdfstring{\HeAlphaHead}{He3-Alpha}: supplementary material\\
\normalsize Interactions and cross sections at next-to-next-to-leading order }

\author{%
 Ratna Khadka\,\orcidlink{0009-0009-6628-7028}}
\email{rk973@.msstate.edu}
\affiliation{Department of Physics \& Astronomy and HPC$^2$ Center for 
Computational Sciences, Mississippi State
University, Mississippi State, MS 39762, USA}

\author{%
Ling Gan\,\orcidlink{0009-0007-0263-7602}}
\email{linggan@arizona.edu Current address: Department of Physics, University of Arizona, Tucson, AZ 85721}
\affiliation{Department of Physics \& Astronomy and HPC$^2$ Center for 
Computational Sciences, Mississippi State
University, Mississippi State, MS 39762, USA}

\author{%
Renato Higa\, \orcidlink{0000-0002-6298-8128}}
\email{higa@if.usp.br}
\affiliation{Instituto de F\'isica, Universidade de S\~ao Paulo, R. do Mat\~ao Nr.1371, 05508-090, S\~ao Paulo, SP, Brazil}

\author{%
Gautam Rupak\,\orcidlink{0000-0001-6683-177X}}
\email{grupak@ccs.msstate.edu}
\affiliation{Department of Physics \& Astronomy and HPC$^2$ Center for 
Computational Sciences, Mississippi State
University, Mississippi State, MS 39762, USA}

\maketitle

We provide additional information relevant to the astrophysical S-factor $S_{34}$ for capture reaction \HeAlpha ~at next-to-next-to-leading order (NNLO) in halo EFT. 

\section{EFT Interactions}
\label{sec:interactions}

Halo EFT~\cite{Higa:2016igc, Premarathna:2019tup} treats the $\alpha$ and $^3$He nuclei as point-like particles interacting with short-ranged nuclear and long-ranged Coulomb forces. The total cross section includes capture to the ground and the first excited state of $^7$Be with spin-parity assignments $\frac{3}{2}^{-}$ and $\frac{1}{2}^{-}$, respectively. Given the $\alpha$  spin-parity $0^+$ and $^3$He spin-parity $\frac{1}{2}^+$, the ground and the first excited state of $^7$Be are considered two-particle bound states $^2P_{3/2}$ and $^2P_{1/2}$, respectively. We use the spectroscopic notation $^{2S+1}L_J$ where $S$ is the total spin, $L$ the total orbital angular momentum, and $J$ the total angular momentum.

The low-momentum capture is dominated by E1 transitions from initial $s$-wave $^2S_{1/2}$ and $d$-wave $^2D_{5/2}$, $^2D_{3/2}$ states. In EFT, initial state $s$-wave strong interaction is considered non-perturbative that appear at leading order (LO) whereas $d$-wave strong interactions are considered perturbative, appearing beyond NNLO, the order of our calculation. Note that $d$-wave with purely Coulomb interactions appear at next-to-leading order (NLO)~\cite{Higa:2016igc,Premarathna:2019tup}. 

The EFT Lagrangian~\cite{Higa:2016igc,Premarathna:2019tup} is divided into several parts for clarity. 
The first is the kinetic energy terms
\begin{multline}
    \mathcal L^{(\text{KE})}=
   \sum_{\alpha=1}^2 {\psi^{\alpha}}^\dagger\left[i\partial_0+\frac{\nabla^2}{2 m_\psi}\right]\psi^{\alpha} 
+\phi^\dagger\left[i\partial_0+\frac{\nabla^2}{2 m_\phi}\right]\phi\\
+\sum_{\zeta,\alpha=1}^2\sum_{i=1}^3{\chi_i^{\alpha,\zeta}}^\dagger\left[\Delta^{(\zeta)}\!+\! \sum_{k=1}^2h_k^{(\zeta)}\left(i\partial_0
+\frac{\nabla^2}{2M}\right)^k\right]\chi_i^{\alpha,\zeta}\,\\
+\sum_{\alpha=1}^2{\chi^{\alpha,s}}^\dagger\left[\Delta^{(s)}
\!+\!\sum_{k=1}^2 h_k^{(s)}\left(i\partial_0
+\frac{\nabla^2}{2M}\right)^k
\right]\chi^{\alpha,s}\,,
\end{multline}
that describe elastic $s$- and $p$-wave scattering. The fermionic $\psi^\alpha$ field with spin index $\alpha$ represents a point-like $^3$He nucleus, the scalar $\phi$ field represents the point-like $\alpha$ nucleus, the auxiliary/dimer field $\chi^{\alpha,\zeta}_i$  with spin index $\alpha$ and vector index $i$ represents the $p$-wave bound states  of $^7$Be, and $\chi^{\alpha,s}$ with spin index $\alpha$ is a dimer field for the $s$-wave $^2S_{1/2}$ state of $\alpha$-$^3$He. $\zeta=0$ is for the $^2P_{3/2}$ ground state and $\zeta=1$ is for the $^2P_{1/2}$ 1st excited state of $^7$Be. The couplings are determined from elastic scattering parameters as below. 

\begin{figure}[tb]
\begin{center}
\includegraphics[width=\columnwidth,clip=true]{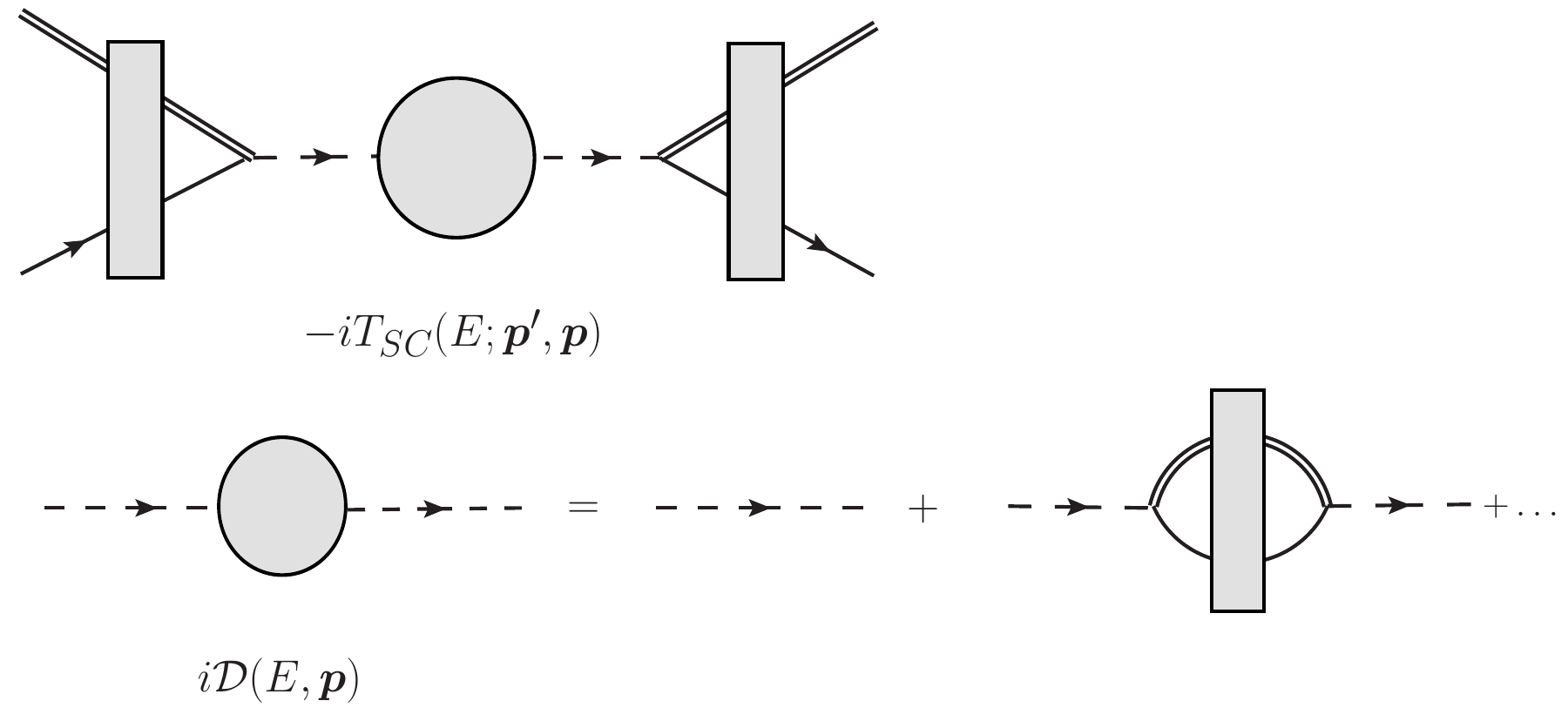}
\end{center}
\caption{\protect  Elastic scattering in $s$- and $p$-wave channels. 
Double lines represent the scalar $\alpha$ particle, single line with an arrow the fermionic $^3$He particle and dashed line the $s$- and $p$-wave dimer as appropriate. The gray rectangles represent long-ranged Coulomb photon exchanges. }
\label{fig:ElasticScattering}
\end{figure}

The $s$-wave dimer propagator from Fig.~\ref{fig:ElasticScattering} is 
\begin{align}
    \begC2{\chi^{\alpha,s}}\endC2{\chi^{\beta,s}}^\dagger 
& =i\mathcal D^{(s)}(p_0,\bm{p})\delta_{\alpha\beta}\,,\nonumber\\
\mathcal D^{(s)}(\frac{p^2}{2\mu},\bm{0})&=\frac{1}{\Delta^{(s)} \!+\sum_{k=1}^2h_k^{(s)} E^k \!-\frac{2\pi}{\mu} J_0(-i p) }\,,
\end{align}
where $J_0(p)$ is the self-energy contribution to the $s$-wave dimer propagator in Fig.~\ref{fig:ElasticScattering}. It is the origin-to-origin Green's function in the presence of Coulomb forces, 
\begin{multline}
J_0(p)=G_C(E;r=0,r'=0)\\
= -2\mu\lambda^{n-3}\int\frac{d^n q}{(2\pi)^n}
\frac{1}{\left[q^2-p^2-i\epsilon\right]}
\frac{2\pi\eta_q}{\left[e^{2\pi\eta_q}-1\right]}\,,
\end{multline}
in $n$ spatial dimensions which is evaluated in dimensional regularization~\cite{Higa:2016igc} as 
\begin{multline}
    J_0(x) =-\frac{\mu}{2\pi}\lambda +\frac{k_C\mu}{2\pi}\left[\frac{1}{n-3}+1-3\gamma_\mathrm{E}\right.\\
\left.+2\ln\frac{\lambda\sqrt{\pi}}{2k_C}-2H(-ik_C/x) \right]\, .
\end{multline}
$\lambda$ is the renormalization group scale and $\gamma_\text{E}\approx0.577216$ is the Euler-Mascheroni constant. Physical observables are independent of $\lambda$.  

The Coulomb-subtracted T-matrix is given by the top Feynman diagram in Fig.~\ref{fig:ElasticScattering}. We match the T-matrix in $s$ wave to the effective range expansion (ERE) 
\begin{align}
    -\mathcal T^{(0)}_\text{SC}(p)&= -\frac{2\pi}{\mu} [C_0(\eta_p)]^2 e^{i2\sigma_0}\mathcal D^{(0)}(E,0)\nonumber\\
    &=\frac{2\pi}{\mu}\frac{e^{i2\sigma_0(p)}}{p\cot\delta_0-ip}\,,
\end{align}
which leads to renormalization conditions
\begin{multline}
    \frac{2\pi}{\mu} J_0(-i p)-\Delta^{(s)} -\sum_{k=1}^2h_k^{(s)} E^k=[C_0(\eta_p)]^2 p(\cot\delta_0-i)\\
    =-\frac{1}{a_0}+\frac{1}{2}r_0 p^2+\frac{1}{4}s_0 p^4 +\cdots -2k_C H(\eta_p)\,,
\end{multline}
and renormalized couplings
\begin{align}
    \Delta^{(s)}&=\frac{1}{a_0}-\lambda+k_C\left[\frac{1}{n-3}+1-3\gamma_\mathrm{E}+\ln\frac{\pi\lambda^2}{4 k_C^2}\right]\,,\nonumber\\
    h_1^{(s)}&=-\mu r_0\,,\nonumber\\
    h_2^{(s)}&=-\mu^2 s_0\,.
\end{align}
This construction can be carried out to higher order in the ERE expansion. However, this is all we need in elastic $s$-wave scattering for the NNLO calculation.

The $p$-wave dimer propagator from Fig.~\ref{fig:ElasticScattering} is 
\begin{align}
\begC2{\chi^{\alpha,\zeta}_i}\endC2{\chi^{\beta,\zeta}_j}^\dagger 
& =i\mathcal D^{(\zeta)}(p_0,\bm{p}) P_{ij}^{(\alpha\beta,\zeta)}\,,\nonumber\\
i\mathcal D^{(\zeta)}(E=\frac{p^2}{2\mu},\bm{0})&=\frac{i}{\Delta^{(\zeta)} +h^{(\zeta)} E -\frac{6\pi}{\mu^3} J_1(-i p)}\,,
\end{align}
where~\cite{Higa:2016igc} 
\begin{align}
    J_1(x)=&-\frac{2\mu\lambda^{3-n}}{n}\int\frac{d^n q}{(2\pi)^n}
\frac{q^2+k_C^2}{\left[q^2-p^2-i\epsilon\right]}
\frac{2\pi\eta_q}{\left[e^{2\pi\eta_q}-1\right]}\nonumber\\
    =&-\frac{\mu k_C}{3\pi}(k_C^2-x^2)H(-i k_C/x)\nonumber\\
    &-\frac{\mu k_c}{3\pi}(k_C^2-x^2) \alpha-\frac{\mu}{3\pi} \beta\, ,\nonumber\\
    \alpha ={}&\frac{1}{n-3}-\frac{4}{3}+\frac{3}{2}\gamma_E-\ln\frac{\lambda\sqrt{\pi}}{2 k_c}+\frac{3\lambda}{4 k_C}\, ,\nonumber\\
    \beta={}& 4\pi^2 k_C^3\zeta'(-2)+\frac{\lambda k_C^2}{4}
    -\frac{3\pi\lambda^2 k_C}{4}+\frac{\pi\lambda^3 }{4}
    \, .
\end{align} 
As in the $s$-wave scattering, we determine the $p$-wave couplings in EFT for capture to final states $\zeta=0$, 1 by matching it to the ERE,
\begin{align}
     -\mathcal T^{(1,\zeta)}_\text{SC}(p) &=-9 [C_1(\eta_p)]^2 e^{i2\sigma_1}\frac{2\pi}{\mu}\frac{p^2}{\mu^2}\mathcal D^{(\zeta)}(E,0)\nonumber\\
     &=\frac{2\pi}{\mu}\frac{e^{2i\sigma_1}}{p\cot\delta_1-i p}\,.
\end{align}
However, we rewrite the ERE as an expansion around the binding momentum $\gamma_\zeta$, writing instead~\cite{Higa:2016igc}
\begin{multline}
    9[C_1(\eta_p)]^2 p^3(\cot\delta_1 -i)=
    2k_C(k_C^2-\gamma_\zeta^2)H(-i\eta_{\gamma_\zeta})\\
    +\frac{1}{2}\rho_1^{(\zeta)}(p^2+\gamma_\zeta^2)
    +\frac{1}{4}\sigma_1^{(\zeta)}(p^2+\gamma_\zeta^2)^2+\dots\\
     -2k_C(k_C^2+p^2)H(\eta_p)\\
     =\mu^2 [\frac{6\pi}{\mu^3} J_1(-i p)-\Delta^{(\zeta)} -h_1^{(\zeta)} E-h_2^{(\zeta)} E^2]\,,
\end{multline}
since the wave function renormalization constant $\mathcal Z^{(\zeta)}$ is the residue of the dimer propagator at the pole at $p=i\gamma_\zeta$. We get the renormalized couplings
\begin{align}
    \mu^2\Delta^{(\zeta)} &= -\frac{1}{2}\rho_1^{(\zeta)}\gamma_\zeta^2-\frac{1}{4}\sigma_1^{(\zeta)}\gamma_\zeta^4\nonumber
    \\&-2k_C(k_C^2-[\gamma^{(\zeta)}]^2)H(-i\eta_{\gamma_\zeta})-2 k_c^3\alpha -2 \beta\nonumber \\
    \mu h_1^{(\zeta)} &= -\rho_1^{(\zeta)} -\sigma_1^{(\zeta)}\gamma_\zeta^2- 4k_C\alpha\,,\nonumber\\
     h_2^{(\zeta)} & =-\sigma_1^{(\zeta)}\,.
\end{align}

The pole in the dimer propagator $\mathcal D^{(\zeta)}(p_0,p)$ for the $p$-wave bound state appears as a zero in the denominator of the propagator. Thus, we write
\begin{multline}\label{eq:wavefunction}
    \frac{1}{\mathcal Z^{(\zeta)}}= \frac{\partial}{\partial E}[\mathcal D^{(\zeta)}(E,0)]^{-1}\Big|_{E=-B_\zeta}\\
    =-\frac{1}{\mu}\left\{\rho_1^{(\zeta)}-4k_C\,H\left(-i\frac{k_C}{\gamma}\right)\right.\\
    \left.
-\frac{2k_C^2}{\gamma^3}(k_C^2-\gamma^2)
\left[\psi'\left(\frac{k_C}{\gamma}\right)
-\frac{\gamma^2}{2k_C^2}-\frac{\gamma}{k_C}\right]\right\}\,.
\end{multline}

\section{EFT Priors}
\label{sec:priors}

In the Bayesian fits, we use the same priors for the $s$-wave scattering parameters as SF3~\cite{Acharya:2024lke}
\begin{align}
    &a_0\sim\mathcal N(\SI{40}{\femto\m},\SI{12}{\femto\m})\,,\nonumber\\ 
    &r_0\sim\mathcal N(\SI{1}{\femto\m},\SI{0.3}{\femto\m})\,,\nonumber\\  
    &s_0\sim\mathcal U(\SI{-30}{\femto\m^3},\SI{30}{\femto\m^3})\,,
\end{align}    
where a large $a_0\sim\Lambda^2/Q^3$ is preferred from the 
power counting~\cite{Higa:2016igc,Premarathna:2019tup} with natural sized $r_0\sim1/\Lambda$ and $s_0\sim1/\Lambda^3$. $Q\sim \SI{70}{\mega\eV}$  and $\Lambda\gtrsim 150$-$\SI{200}{\mega\eV}$ are the estimated low momentum and cutoff momentum scales, respective, of the EFT in the $Q/\Lambda$ expansion. The choice of priors here and below were informed by previous EFT~\cite{Higa:2016igc,Premarathna:2019tup} and SONIK~\cite{Paneru:2024} analyses. The fits are insensitive to the choice of uniform $\mathcal U$ vs normal $\mathcal N$ priors. We kept the priors in this analysis consistent with SF3~\cite{Acharya:2024lke}. 

For the $p$-wave scattering parameters we use uniform priors~\cite{Acharya:2024lke}
\begin{align}
    &\rho_1^{(0)}\sim\mathcal U(\SI{-300}{\mega\eV},\SI{-48}{\mega\eV})\,,\nonumber\\
    &\sigma_1^{(0)}\sim\mathcal U(\SI{-5}{\femto\m},\SI{5}{\femto\m})\,,\nonumber\\
    &\rho_1^{(1)}\sim \mathcal U(\SI{-300}{\mega\eV},\SI{-33}{\mega\eV})\,,\nonumber\\
    &\sigma_1^{(1)}\sim \mathcal U(\SI{-5}{\femto\m},\SI{5}{\femto\m})\,,
    \end{align}
    where the upper bound on the effective momenta $\rho_1^{(\zeta)}\sim Q$ was constrained by the physical requirement that the wave function renormalization constant $\mathcal Z^{(\zeta)}>0$. The shape parameters $\sigma_1^{(\zeta)}\sim 1/\Lambda$ do not contribute to the capture but they appear in the $p$-wave phase shift. The systematic error 
in the phase shift data is not well documented~\cite{Spiger:1967,Boykin:1972}.  We accommodate unknown errors in the phase shift data by inflating the reported uncertainty $\sigma^2\rightarrow \sigma^2+K^2$ with $K\sim\mathcal U(0^\circ,10^\circ)$ drawn from a uniform distribution~\cite{Higa:2016igc,Premarathna:2019tup} consistent with the fitting uncertainty reported in Ref.~\cite{Boykin:1972}. At NLO and NNLO, 
the fits give $K=\ang{0.28\pm0.21}$ and $K=\ang{0.29\pm0.22}$, respectively,  and the uncertainty $K$ is not expected to play an outsized roles in the fits.  

The two-body current couplings are assumed to be natural. We use priors: 
\begin{align}
&L_\text{E1}^{(0)}\sim \mathcal N(1,0.5)\,,& &L_\text{E1}^{(1)}\sim \mathcal N(1,0.5)\,,\nonumber\\ 
&L_\text{E1}^{(0,s)}\sim \mathcal N(0,\SI{0.5}{\femto\m^2})\,,
& &L_\text{E1}^{(1,s)}\sim \mathcal N(0,\SI{0.5}{\femto\m^2})\,,\nonumber\\ 
&L_\text{E1}^{(0,p)}\sim \mathcal N(0,\SI{0.5}{\femto\m^2})\,,
& &L_\text{E1}^{(1,p)}\sim \mathcal N(0,\SI{0.5}{\femto\m^2})\,.
\end{align}
The fits find non-zero values for the LO two-body current couplings $L_\text{E1}^{(\zeta)}$ of either sign with a slightly higher evidence for positive values~\cite{Higa:2016igc,Premarathna:2019tup} that we build into the choice of priors. For the NNLO two-body current couplings $L_\text{E1}^{(s,\zeta)}$, $L_\text{E1}^{(p, \zeta)}$ we pick priors of size $1/\Lambda^2$ of either sign.

\section{S-factor and Branching ratio}
\label{sec:sfactor}

In this section we provide numerical values for the S-factor $S_{34}$  and branching ratio $R$  at LO, NLO and NNLO for various center-of-mass (cm) energy $E$ in Table~\ref{table:S34RatioBayesian}. We provide both the errors from Bayesian fits and the estimated EFT uncertainty from higher orders contributions. 

The analytic form for the S-factor provided in the main text was fitted to the numerical NNLO values at $E/\si{\kilo\eV}=60\times10^{-3}$, 0.5, 1, 11, 21, 23, 31, 41, 50, 100, 200, 500, 700, 1000, 1200, 1500, 1600.

\begin{table*}[htb]
\centering
\caption{\protect \HeAlpha ~astrophysical S-factor $S_{34}$ and 
branching ratio $R$ for capture to the 1st excited relative to the ground state of $^7$Be as a function of cm energy $E$. The mean values of the Bayesian fit to data as described in the main text are presented. The first set of errors are the standard deviations from the Bayesian fits. The second set is the estimated LO 30\%, next-to-leading order 10\%, NNLO 3\%  EFT errors, respectively, from higher order corrections. }
\begin{ruledtabular}
\begin{tabular}{rllllll}
$E$ (\si{\kilo\eV}) & $S_{34}^\text{(LO)}$ (\si{\kilo\eV\barn})
& $S_{34}^\text{(NLO)}$ (\si{\kilo\eV\barn}) & $S_{34}^\text{(NNLO)}$ (\si{\kilo\eV\barn})
& $R^\text{(LO)}$ 
& $R^\text{(NLO)}$  & $R^\text{(NNLO)}$ 
\\  \hline \rule{0pt}{0.9\normalbaselineskip}
\begin{filecontents*}{S34-Ratio-list-Bayesian-2000keV-2025-9-11.csv}
energy,LO,dLO,dEFTLO,NLO,dNLO,dEFTNLO,NNLO,dNNLO,dEFTNNLO,RLO,dRLO,dREFTLO,RNLO,dRNLO,dREFTNLO,RNNLO,dRNNLO,dREFTNNLO
0.06,0.557,17,167,0.561,17,56,0.564,16,17,0.399,13,120,0.395,9,39,0.413,20,12
0.5,0.557,17,167,0.561,17,56,0.564,16,17,0.399,13,120,0.395,9,39,0.413,20,12
1,0.557,17,167,0.561,17,56,0.564,16,17,0.399,13,120,0.395,9,39,0.413,20,12
11,0.555,17,166,0.557,17,56,0.561,16,17,0.399,12,120,0.395,8,39,0.413,20,12
21,0.552,16,166,0.554,16,55,0.558,15,17,0.399,12,120,0.395,8,39,0.412,19,12
23,0.551,16,165,0.553,16,55,0.557,15,17,0.399,12,120,0.395,8,39,0.412,19,12
31,0.549,16,165,0.551,16,55,0.554,15,17,0.399,12,120,0.394,8,39,0.411,19,12
41,0.547,15,164,0.548,15,55,0.551,14,17,0.399,12,120,0.394,8,39,0.411,18,12
50,0.544,15,163,0.545,15,54,0.548,14,16,0.399,11,120,0.394,8,39,0.410,18,12
100,0.531,13,159,0.529,13,53,0.532,12,16,0.400,10,120,0.394,7,39,0.407,15,12
150,0.517,12,155,0.513,11,51,0.516,11,15,0.400,9,120,0.394,6,39,0.405,13,12
200,0.503,11,151,0.497,10,50,0.500,10,15,0.400,8,120,0.395,5,39,0.403,10,12
250,0.488,9,146,0.481,9,48,0.483,9,14,0.401,8,120,0.395,5,40,0.401,9,12
300,0.473,9,142,0.465,8,47,0.466,8,14,0.401,7,120,0.396,4,40,0.400,7,12
350,0.458,8,137,0.449,7,45,0.450,7,14,0.402,6,121,0.397,4,40,0.400,6,12
400,0.442,8,133,0.434,7,43,0.434,7,13,0.403,5,121,0.397,3,40,0.400,5,12
450,0.428,7,128,0.420,7,42,0.419,6,13,0.404,5,121,0.399,3,40,0.400,4,12
500,0.414,7,124,0.406,6,41,0.405,6,12,0.404,4,121,0.400,3,40,0.400,4,12
550,0.400,7,120,0.394,6,39,0.393,6,12,0.405,4,122,0.401,3,40,0.401,4,12
600,0.388,7,116,0.383,6,38,0.381,6,11,0.406,3,122,0.403,3,40,0.403,4,12
650,0.376,6,113,0.372,6,37,0.371,6,11,0.407,3,122,0.404,3,40,0.404,4,12
700,0.365,6,110,0.363,5,36,0.361,5,11,0.407,3,122,0.406,3,41,0.406,4,12
750,0.355,6,107,0.355,5,35,0.353,5,11,0.408,4,122,0.408,3,41,0.408,3,12
800,0.346,6,104,0.347,5,35,0.346,5,10,0.409,4,123,0.410,3,41,0.410,3,12
850,0.338,6,101,0.341,5,34,0.340,5,10,0.410,4,123,0.412,3,41,0.412,3,12
900,0.330,6,99,0.336,5,34,0.334,5,10,0.411,5,123,0.414,3,41,0.414,3,12
950,0.323,6,97,0.331,5,33,0.330,5,10,0.412,5,124,0.416,3,42,0.416,3,12
1000,0.317,7,95,0.327,5,33,0.326,5,10,0.413,6,124,0.419,3,42,0.419,3,13
1100,0.306,7,92,0.321,5,32,0.320,5,10,0.414,7,124,0.423,3,42,0.424,3,13
1200,0.297,7,89,0.318,5,32,0.317,5,10,0.416,8,125,0.428,3,43,0.428,3,13
1300,0.289,8,87,0.316,5,32,0.315,5,9,0.418,9,125,0.433,3,43,0.433,4,13
1400,0.283,9,85,0.316,5,32,0.315,5,9,0.419,10,126,0.439,5,44,0.436,5,13
1500,0.278,9,83,0.318,5,32,0.317,5,10,0.421,11,126,0.444,6,44,0.439,6,13
1600,0.274,10,82,0.321,6,32,0.319,5,10,0.422,12,127,0.449,8,45,0.442,7,13
1700,0.271,11,81,0.325,6,32,0.323,6,10,0.424,13,127,0.455,11,45,0.443,9,13
1800,0.268,11,80,0.330,6,33,0.327,6,10,0.425,14,128,0.460,14,46,0.444,12,13
1900,0.266,12,80,0.336,6,34,0.332,6,10,0.427,15,128,0.466,17,47,0.443,15,13
2000,0.265,13,79,0.342,6,34,0.338,6,10,0.428,16,129,0.471,21,47,0.442,20,13
2100,0.264,14,79,0.349,7,35,0.345,7,10,0.430,17,129,0.477,25,48,0.439,25,13
2200,0.263,14,79,0.357,7,36,0.352,7,11,0.431,18,129,0.482,29,48,0.435,32,13
2300,0.263,15,79,0.365,8,37,0.359,8,11,0.432,19,130,0.488,34,49,0.431,40,13
2400,0.263,16,79,0.374,8,37,0.367,8,11,0.434,19,130,0.493,38,49,0.425,50,13
2500,0.263,17,79,0.383,9,38,0.376,9,11,0.435,20,130,0.498,44,50,0.417,61,13
2600,0.264,18,79,0.393,9,39,0.385,10,12,0.436,21,131,0.503,49,50,0.409,73,12
2700,0.265,19,79,0.403,10,40,0.394,11,12,0.437,22,131,0.508,55,51,0.400,87,12
2800,0.266,19,80,0.413,11,41,0.403,12,12,0.439,22,132,0.514,61,51,0.389,103,12
\end{filecontents*}
\csvreader[head to column names, late after line=\\]{S34-Ratio-list-Bayesian-2000keV-2025-9-11.csv}{}
{ \energy 
& \LO(\dLO)(\dEFTLO)
& \NLO(\dNLO)(\dEFTNLO)
& \NNLO(\dNNLO)(\dEFTNNLO)
& \RLO(\dRLO)(\dREFTLO)
& \RNLO(\dRNLO)(\dREFTNLO)
& \RNNLO(\dRNNLO)(\dREFTNNLO)
}
\end{tabular}
\end{ruledtabular}
 \label{table:S34RatioBayesian}
\end{table*}

%